\documentclass[a4paper]{article}
\usepackage{jheppub}

\usepackage[familydefault]{Rosario}
\usepackage[utf8]{inputenc}
\usepackage{graphicx}
 \graphicspath{{./figures/}}
\usepackage{dcolumn} 
\usepackage{amsmath,amsfonts} 
\allowdisplaybreaks

\usepackage{mathabx}
\usepackage{physics}
\usepackage{slashed}
\usepackage{bm}
\usepackage{cancel}
\usepackage[caption=false]{subfig}
\usepackage{wrapfig}
\usepackage[dvipsnames,svgnames]{xcolor}
\usepackage{braket}
\usepackage{multirow}
\usepackage{graphicx}
\usepackage{hhline}
\usepackage[shortlabels]{enumitem}
\usepackage{float}
\newenvironment{changemargin}[2]{%
\begin{list}{}{%
\setlength{\topsep}{0pt}%
\setlength{\leftmargin}{#1}%
\setlength{\rightmargin}{#2}%
\setlength{\listparindent}{\parindent}%
\setlength{\itemindent}{\parindent}%
\setlength{\parsep}{\parskip}%
}%
\item[]}{\end{list}}

\usepackage[capitalise]{cleveref}

\hypersetup{%
  colorlinks=true, linktocpage=true, pdfstartpage=1, pdfstartview=FitV,%
  breaklinks=true, pdfpagemode=UseNone, pageanchor=true, pdfpagemode=UseOutlines,%
  plainpages=false, bookmarksnumbered, bookmarksopen=true, bookmarksopenlevel=1,%
  hypertexnames=true, pdfhighlight=/O,
  urlcolor=Cerulean, linkcolor=Cerulean, citecolor=Cerulean, 
}

\usepackage{framed}
\colorlet{shadecolor}{blue!20}
\usepackage[normalem]{ulem}

\begin{document}

\title{Oscillating Resonances: Imprints of ultralight dark matter at colliders}

\author[]{Martin Bauer and Sreemanti Chakraborti}

\affiliation[]{Institute for Particle Physics Phenomenology, Department of Physics,\\Durham University,\\Durham, UK}

\emailAdd{martin.m.bauer@durham.ac.uk}
\emailAdd{sreemanti.chakraborti@durham.ac.uk}

\preprint{IPPP/26/08}

\abstract{
In models where ultralight fields constitute dark matter, the misalignment mechanism leads to coherent, low-amplitude oscillations in fundamental constants. This effect arises from effective operators that couple dark matter to Standard Model fields. We present different models that can induce these effective operators by integrating out a mediator field. For mediator masses within the reach of collider searches, an alternative way to discover ultralight dark matter is to search for a resonance. Due to being a mediator to dark matter, the mass of the mediator oscillates. The resonance therefore, should not appear as a single isolated peak, but is smeared out once data is averaged over an oscillation cycle or more. Remarkably, the oscillation period and amplitude are in the range of current and future collider searches, even though constraints from atomic clocks probing variations of the fine-structure constant and the electron mass are very strong. We recast existing searches and projections from Belle II, LHCb and SHiP for an `oscillating resonance', and discuss how the periodicity of the signal can be used to reconstruct the peak from mass-binned data. We further show that time-stamped data would allow to unfold the signal via a fast Fourier transform and determine the significance of the signal for different background levels. The discovery of an oscillating resonance at a collider, with characteristics as predicted in ultralight dark matter scenarios, would constitute a powerful probe of dark matter's underlying nature. } 

\maketitle


\section{Introduction}

Despite evidence for dark matter (DM) at astrophysical and cosmological scales, a successful detection of non-gravitational interactions between dark matter and Standard Model (SM) particles continues to elude us. One possibility is that dark matter is too light to produce signals above the detection threshold of direct detection experiments and indirect detection experiments. Ultralight bosonic dark matter, with masses well below an eV, can account for the observed relic abundance via the misalignment mechanism and provides an alternative to weakly interacting massive particles~\cite{Hu:2000ke, Schive:2014dra, Hui:2016ltb}. In contrast to heavier, particle-like dark matter candidates, ultralight dark matter (ULDM) can be treated as a coherently oscillating classical background field. Interactions of ultralight dark matter with photons and electrons induce oscillations of fundamental constants such as the fine-structure constant or the electron mass~\cite{Olive:2001vz, Stadnik:2014tta, Safronova:2017xyt, VanTilburg:2015oza}. In most phenomenological studies, these interactions are mediated by higher-order operators in an effective field theory connecting the dark matter field with Standard Model particles. Strong constraints on these Wilson coefficients arise from measurements of transition frequencies in atomic clocks~\cite{ Arvanitaki:2014faa, Stadnik:2016zkf, Kennedy:2020bac}, from big bang nucleosynthesis~\cite{Blum:2014vsa,  Sibiryakov:2020eir} and from tests of the equivalence principle \cite{Touboul:2017grn, Touboul:2022yrw}. Here, we consider the consequences of an explicit mediator for different ultraviolet (UV) completions of these effective operators. We discuss the properties of vector and scalar mediators that interact quadratically with the dark matter field.  In these models, dark matter does not couple to any SM particle at the tree level. However,  dark matter couplings to electrons and photons occur at one and two loop, respectively. If the mass of the mediator is in reach of colliders it can be produced resonantly and detected via decays into SM particles. Due to the direct couplings to ULDM, the mediator mass and the mediator couplings oscillate. The timescale of these oscillations is directly linked to the dark matter mass and it is remarkable that the time scales that can be probed at colliders coincide with the interesting dark matter mass range $10^{-22}$ eV $\lesssim m_\phi \lesssim 10^{-15}$ eV. 

Instead of a narrow resonance peak, for an `oscillating resonance', the signal at a collider would be suppressed and smeared over several mass bins arising from averaging events over timescales comparable to, or longer than, the oscillation period. At the example of the limits on $e^+ e^-\to  S e^+ e^- \to \ell^+\ell^- e^+ e^-$ from Belle~\cite{Cogollo:2024fmq}, projections for resonance searches in $pp \to X \to \mu^+\mu^-$ at LHCb run 6~\cite{Craik:2022riw} and $e^+ e^-\to S e^+ e^- \to \ell^+\ell^- e^+ e^-$ and $e^+ e^-\to X \gamma \to \ell^+\ell^- \gamma$ at Belle II~\cite{Ferber:2015jzj, Belle-II:2018jsg}, we demonstrate how far constraints on mediator interactions to SM particles are weakened if interpreted as searches for mediators interacting with an oscillating background field. For long-lived mediators, the dependence on the oscillating parameters may effectively cancel, and we discuss the implications for different UV completions at the example of projections for hidden photon searches at SHiP~\cite{SHiP:2021nfo}. We compare the collider reach for such mediators with constraints from variations of fundamental constants. Intriguingly, the parameter space that can be probed at collider searches is not excluded by constraints from atomic clocks or tests of the equivalence principle. 

The recast bump-hunt searches are not expected to provide optimal sensitivity to an `oscillating resonance'. One of the main results of this paper is that incorporating the signal’s periodic time dependence can improve collider sensitivity to mediators coupled to ultralight dark matter. For mass-binned analyses, we present a peak-finding or peak-merging algorithm that reconstructs the central peak, provided that the signals at both edges of the modulation window can be isolated. This approach can be applied to recast existing limits, but it is limited and the algorithm cannot recover the total event yield. However, this information is not lost and can be recovered using time-binned data. We show how a fast Fourier transform can reconstruct the resonance peak in a sample of simulated, time-binned data. For the analysis of time-binned data we take into account different background levels and determine the signal-to-noise ratio for which the fast Fourier transform cannot reliably find the signal.   
  
The paper is structured as follows: Section \ref{sec:model} introduces different mediator models and discusses their properties. In Section \ref{sec:DMcouplings} we compute the loop-induced coefficients of the effective dark matter interactions with photons and electrons and compute the strongest constraints from oscillating fine-structure constant. In Section \ref{sec:collider} we present the reinterpreted sensitivity reach of existing and future collider bounds  and in Section \ref{sec:reconstruct} we introduce a reconstruction technique to recover sensitivity to oscillating resonances using mass-binned and time-stamped data. We conclude in Section \ref{sec:conclusions}.


\section{A dark mediator in ultralight dark matter background}\label{sec:model}

Interactions between the Standard Model and ultralight dark matter, here a real scalar field $\phi$,  are usually represented by effective interactions with photons, electrons or nucleons as appropriate at low energy scales. We go a step beyond and introduce mediators between standard model particles and dark matter. 
We consider two different scenarios, different by the way the dark matter scalar interacts with the Standard Model. In the first scenario, a dark scalar $S$ couples directly to the dark matter as well as Standard Model fermions
\begin{align}\label{eq:LS}
\mathcal{L_S}=  \partial_\mu S \partial^\mu S^\dagger -\mu_S^2S^\dagger S -d_S\frac{\mu_S^2}{\Lambda} \phi S^\dagger S + \lambda_S (S^\dagger S)^2+S\sum_f \kappa_f \bar f f\,,
\end{align}
where $m_S^2>0$ and the scalar $S$ has no vacuum expectation value. We assume a potential Higgs portal term with both $\phi$ and $S$ to be negligibly small. 
As a result of the dark matter interactions, the scalar has a $\phi-$dependent mass
\begin{align}
m_S^2(\phi) =\mu_S^2\left(1+\frac{d_S}{\Lambda}\phi\right)\,.
\end{align}
In the second scenario, we assume a dark gauge boson $X$ interacting with SM fermions through kinetic mixing with the SM photon or via gauge couplings to SM fermions. The gauge boson obtains its mass through interactions with a scalar $H_D$, which is charged under the hidden gauge symmetry $U(1)_D$. The new physics Lagrangian is then given by
\begin{align}
    \mathcal{L}_X&=-\frac{1}{4}X_{\mu\nu}X^{\mu\nu}+\frac{1}{2}m_X^2X_\mu X^\mu+g_X X^\mu \sum_f Q_f \bar{f} \gamma_\mu f\nonumber-\frac{\epsilon^0}{2}\,F_{\mu\nu}X^{\mu\nu}\\
    &+\frac{1}{2}|\partial_\mu\phi|^2+|D_\mu H_D|^2-V(H_D)\,,
    \label{eq:gauge}
\end{align}
where $D_\mu=\partial_\mu+i\, g_D X_\mu$ and $m_X=g_X\, v_D$, $v_D$ being the dark vev. For a dark photon which interacts with the SM through kinetic mixing, for which one can replace  $g_X=\epsilon e$. Here, we can distinguish three different interactions between the dark matter field by adding the operators
\begin{align}
\mathcal{L}_1&= \frac{\mu_D^2 d_H}{\Lambda}\phi|H_D|^2\,,\label{eq:dim3}\\
\mathcal{L}_2&= -c_\phi\frac{\phi}{\Lambda}X_{\mu\nu}X^{\mu\nu}\,,\label{eq:dim5}\\
\mathcal{L}_3&= -\epsilon^0\tilde c_\phi\frac{\phi}{\Lambda}X_{\mu\nu}F^{\mu\nu}\,\label{eq:dim5eps}.
\end{align}
For the case of $\mathcal{L}_1$, the mass term in the dark Higgs potential is modified,
\begin{align}
    V(H_D)-\mathcal{L}_1 &= -\mu_D^2|H_D|^2-\,  \frac{\mu_D^2 d_H}{\Lambda}\phi|H_D|^2+\lambda_D 
    |H_D|^4\nonumber\\
    &= -\mu_D^2\left(1 + d_H\frac{\phi}{\Lambda}\right) |H_D|^2+\lambda_D 
    |H_D|^4\,,
    \label{eq:scalar}
\end{align}
leading to a field-dependent vacuum expectation value 
\begin{align}
    v_D(\phi)=v_D^0\left(1+d_H\frac{\phi}{2\Lambda}\right)\,,
\end{align}
where $v_D^0=\mu_D/\sqrt{\lambda_D}$.
In the case of $\mathcal{L}_2$, the gauge coupling of the $X-$boson is field-dependent,
\begin{align}
g_X(\phi)=g_X^0\left(1+c_\phi\frac{\phi}{\Lambda}\right)\,,
\end{align}
where $g^0_X$ is the gauge coupling in the limit $\Lambda \to \infty$. 
In the case of $\mathcal{L}_3$ the kinetic mixing parameter is field dependent,
\begin{align}
\epsilon(\phi)=\frac{\epsilon^0}{2}\left(1+\tilde c_{\phi}\frac {\phi}{\Lambda}\right)\,.
\label{eq:osc_eps}
\end{align}
In general, a combination of the parameters $d_H, c_\phi, \tilde c_\phi$ can be non-zero, but we will focus on the different phenomenology for the cases where only one of the coefficients is non-zero.

In the ultralight mass regime, bosonic dark matter behaves like a plane wave with the time period of oscillation determined by its Compton mass and the amplitude dictated by the dark matter energy density ($\rho_{\rm DM}$). The plane-wave solution for the ULDM field is 
\begin{align}
    \phi(t)=\frac{\sqrt{2\rho_{\rm DM}}}{m_\phi}\cos\omega t\,,
\end{align}
where $\omega$ denotes the Compton frequency of ULDM and the amplitude is $\phi_0=\sqrt{2\rho_{\rm DM}}/{m_\phi}$. As a result, for $\mathcal{L}_S, \mathcal{L}_1$ and $\mathcal{L}_2$, the scalar $S$ or the $X$ boson have a time-dependent mass, respectively 
\begin{align}
m_S(t)&=m_S^0\left(1+\frac{\delta m_S}{m_S^0}\cos\omega t\right)\,,\\
m_{X}(t) &= g_X\, v_D(t) =g_X(t)\, v_D \equiv m_X^0\left(1+\frac{\delta m_X}{m_X^0}\cos\omega t\right)\,,
\label{Eq:massmod}
\end{align}
where
\begin{align}
\delta m_S&=\frac{1}{2}d_S\frac{\phi_0}{\Lambda}m_S^0\qquad \text{for}\quad \mathcal{L}_S\,,\label{eq:LSL2L3delta1}\\
\delta m_X& =d_H\frac{\phi_0}{2\Lambda} m_X^0\qquad \text{for}\quad \mathcal{L}_1\,,\label{eq:LSL2L3delta2}\\
\delta m_X& =c_\phi\frac{\phi_0}{\Lambda} m_X^0\qquad \text{for}\quad \mathcal{L}_2\,.\label{eq:LSL2L3delta3}
\end{align}
In the case of $\mathcal{L}_3$, the mass of the X boson is time-dependent due to the corrections to the mass eigenvalue after rotating into the 
mass eigenbasis
\begin{align}
m_X^2(t)=(m_X^0)^2\left(1+\epsilon(t)^2(s_W^2+\delta ) \right)\,,
\end{align}
where $\delta=(m_X^0)^2/M_Z^2$~\cite{Bauer:2018onh}. The same effect also induces an oscillation of the Z boson mass, and we will therefore consider this scenario in a future work. The scalar scenario $\mathcal{L}_S$ and the interactions described by $\mathcal{L}_1$ have in common that both lead to field-dependent mediator masses, whereas the model described by $\mathcal{L}_2$ introduced a field-dependent mediator mass as well as field-dependent gauge couplings. In the following, we will focus on the phenomenological implications of these two cases. 

\section{Dark Matter coupling to photons and electrons}
\label{sec:DMcouplings}

\begin{figure}
\centering
\includegraphics[width=.9\textwidth]{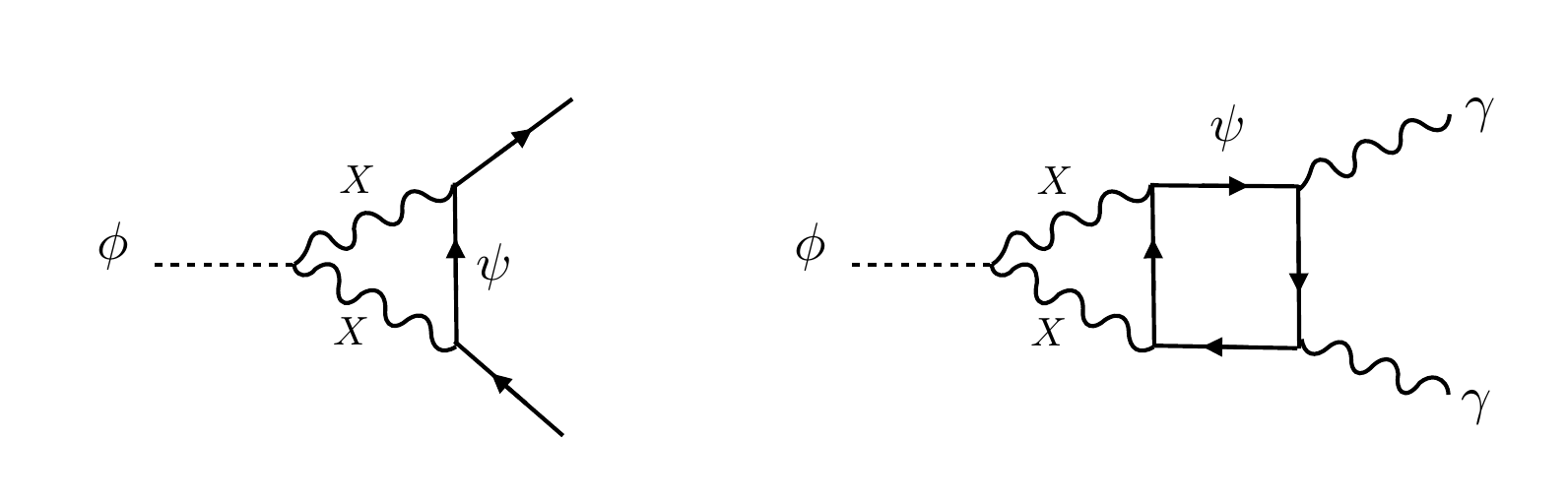}
\caption{Leading-order Feynman diagrams that induce dark matter couplings to SM fermions and photons for a vector mediator $X$. The analogous diagrams induce couplings to photons and fermions in the scalar mediator model by replacing $X$ by $S$. \label{fig:diagrams}}
 \end{figure}

The strongest constraints on ultralight dark matter arise from limits on the variation of the electron mass and the fine-structure constant~\cite{Olive:2001vz, Stadnik:2014tta, Safronova:2017xyt, VanTilburg:2015oza}. These variations are induced by the operators
\begin{align}
\mathcal{L}_\text{eff}=-m_e \, c_\psi\, \phi \,\bar \psi\psi -\frac{1}{4}c_\gamma\, \phi F_{\mu\nu}F^{\mu\nu}\,.
\end{align}
In the scenarios discussed here,  couplings to electrons and photons are induced at one-loop and two-loop, respectively, as shown in Fig.~\ref{fig:diagrams}.
In the case of a scalar mediator, the effective interactions with electrons read
\begin{align}\label{eq:Scalar}
c_\psi^S=\frac{\kappa_e^2}{8\pi^2} d_S\frac{1}{\Lambda}\frac{\mu_S^2}{m_S^2}I_S&=\frac{\kappa_e^2}{4\pi^2} d_S\frac{1}{\Lambda}\frac{\mu_S^2}{m_S^2}\left(\frac32+\tau\left(\frac76+\log\tau\right)+\mathcal{O}(\tau^2)\right)\,
\end{align}
where $\tau=m_e^2/M_S^2$
in the limit $m_\phi \ll m_\psi< m_X$. In the case of a vector mediator $\mathcal{L}_1$ with gauge couplings to SM fermions $g_X$, the coupling to the $X$ bosons is induced via the mixing with the dark higgs boson, which is a renormalisable operator and these loops are also finite. We find in agreement with~\cite{Batell:2009yf},
\begin{align}\label{eq:Xmass}
c_\psi^X=\frac{g_X^2}{4\pi^2}\frac{d_H}{\Lambda}\frac{v_D^2}{m_X^2}I_X=\frac{g_X^2}{4\pi^2}\frac{d_H}{\Lambda}\frac{v_D^2}{m_X^2}\left(\frac32+2\tau\left(\frac53+\log\tau\right)+\mathcal{O}(\tau^2)\right)\,,
\end{align}
where $\tau_X= m^2/m_X^2$. 
If the case of dark matter interacting with the field-strength tensor of $U(1)_X$ as in $\mathcal{L}_2$, the loop is divergent and the effective coefficient reads
\begin{align}\label{eq:XFS}
c_\psi^\text{FS}(\mu)=\frac{g_X^2}{8\pi^2}c_\phi\frac{1}{\Lambda}\left(I_\text{FS}-\frac{5}{2}\right)=\frac{g_X^2}{8\pi^2}\frac{1}{\Lambda}c_\phi\left(\frac12+3\log\frac{\mu^2}{m_X^2}+4\tau+\mathcal{O}(\tau^2)\right)\,.
\end{align}
The effective coefficients in the models in which the interaction is mediated by a $U(1)_X$ gauge boson both assume that SM fermions are charged under $U(1)_X$. As a result, the interaction strength with electrons is directly proportional to the variation of the mass of the $U(1)_X$ gauge boson. To leading order in $\tau_X$ one can write
\begin{align}\label{eq:wilson}
c_\psi^X &=\frac{3}{4\pi^2}\frac{\delta m_X}{m_X}\frac{\, m_\phi}
{\sqrt{2\rho_\text{DM}}} \\
c_\psi^\text{FS}(\mu)&=\frac{g_X^2}{8\pi^2}\frac{\delta m_X}{m_X}\frac{\, m_\phi}
{\sqrt{2\rho_\text{DM}}}\left(1+6\log\frac{\mu^2}{m_X^2}\right)
\end{align}

In the case of dark matter mixing with the dark Higgs, the interaction with electrons is a function of the variation of the mass of the $U(1)_X$ mediator and the dark matter mass. For dark matter interactions with the field strength tensor of $U(1)_X$ instead there is also a factor of the gauge coupling $g_X^2$. If interactions with the SM fermions are induced via kinetic mixing of the $U(1)_X$ boson with the photon, one instead replaces $g_X^2\to  4\pi \alpha \epsilon^2$, so that the Wilson coefficients in \eqref{eq:wilson} are rescaled by $\epsilon^2\alpha/\alpha_X$ instead. In this case the interaction with fermions is suppressed for large $g_X$ and small $\epsilon$. Similarly, in the case of the scalar mediator, we find
\begin{align}
c_\psi^S=\frac{3\kappa_e^2}{4\pi^2} \frac{\delta m_S}{m_S}\frac{\, m_\phi}
{\sqrt{2\rho_\text{DM}}}\frac{\mu_S^2}{m_S^2}\,,
\end{align}
where the coupling to fermions and the scalar mass are independent parameters. The $\phi-$dependence of the fermion mass can then be written as 
\begin{align}\label{eq:mpsieff}
 m_\psi^\text{eff}(\phi)=m_\psi +\delta m_\psi=m_\psi(1+c_\psi\phi)\,,   
\end{align}
where $c_\psi=c^S_\psi, c^X_\psi$ or $c_\psi^\text{FS}$ depending on the underlying UV completion. The variation of the electron mass can then be written as 
\begin{align}
\frac{\delta m_e}{m_e} = \begin{cases}\displaystyle\frac{3}{4\pi^2} \frac{\delta m_X}{m_X}\cos(\omega t)\,,& c_\psi=c_\psi^X,\\\displaystyle
\frac{g_X^2}{8\pi^2} \frac{\delta m_X}{m_X}\cos(\omega t)\,,& c_\psi=c_\psi^\text{FS},\\\displaystyle
\frac{3\kappa_e^2}{4\pi^2} \frac{\delta m_S}{m_S}\frac{\mu_S^2}{m_S^2}\cos(\omega t)\,,& c_\psi=c_\psi^S,
\end{cases}
\end{align}
establishing a direct connection between the variation of the electron mass, the free parameters of the model and the amplitude of the variation of the mediator mass.

At the two-loop level, the dark matter coupling to the mediator such as a scalar or $U(1)_X$ boson induces a dark matter coupling to the electromagnetic field strength tensor. For $m_X, m_S > m_e$, the loop diagram corresponds to the top-loop contribution to the Higgs diphoton coupling with an effective fermion interaction given by \eqref{eq:scalar}, \eqref{eq:Xmass} and \eqref{eq:XFS}.

\begin{figure}[h!]
\begin{changemargin}{-2.3cm}{-1.5cm}
\centering\includegraphics[width=.63\textwidth]{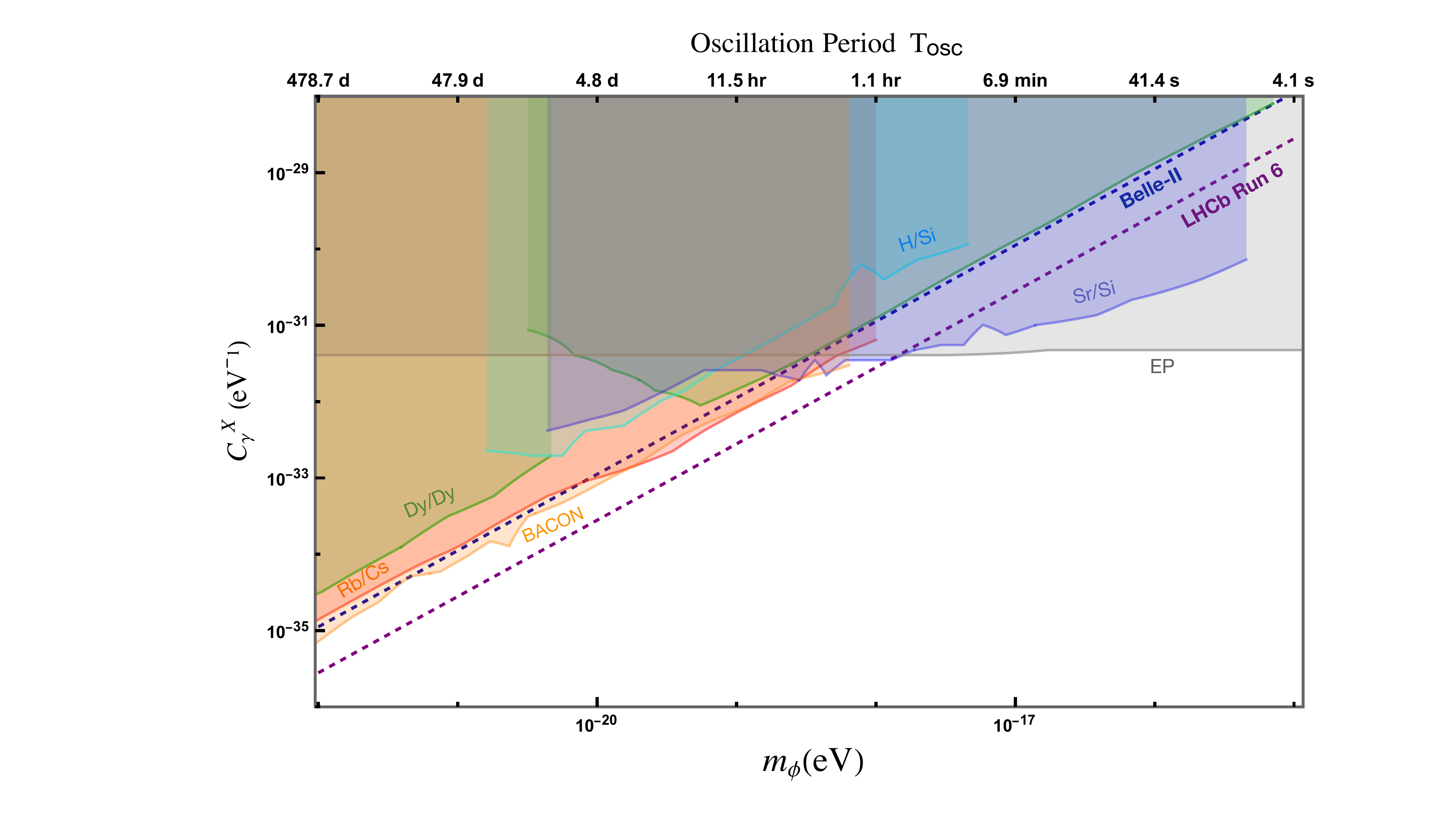}
\hspace{-1cm}\raisebox{.15cm}{\includegraphics[width=.61\textwidth]{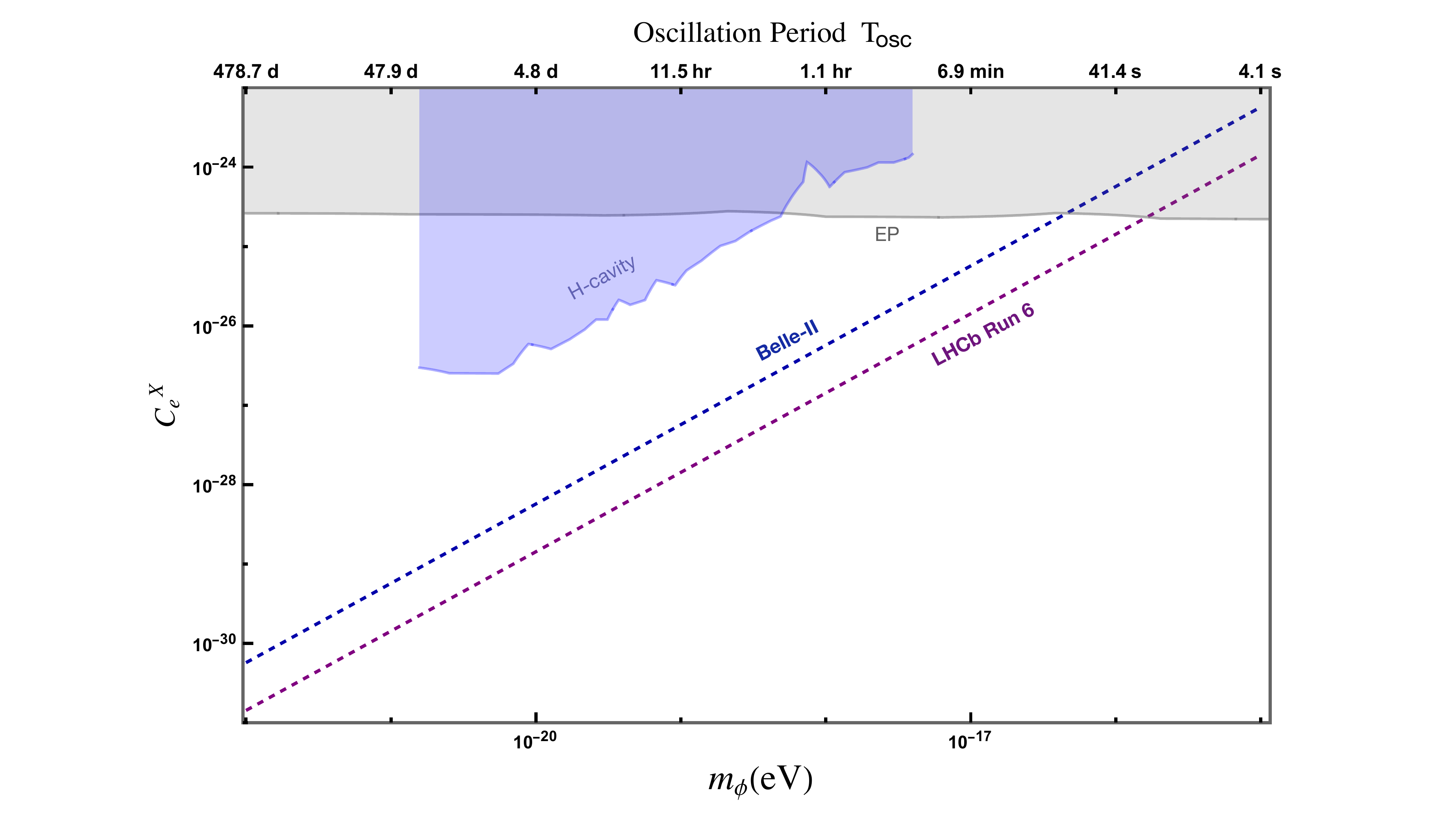}}
\end{changemargin}
    \caption{The loop-induced photon and electron couplings for a dark photon $X$ in the ULDM background. The collider bounds (LHCb and Belle-II) correspond to a 10\% mass modulation and $g_X = 4\pi$. The shaded lines denote the latest bounds from precision sensors such as atomic clocks (Rb/Cs, BACON), atomic spectroscopy (Dy/Dy), clock-cavity comparisons (H/Si, Sr/Si) and EP violation experiment (MICROSCOPE).}
    \label{fig:loopmodel1}
\end{figure}
\begin{figure}[h!]
 \begin{changemargin}{-2.3cm}{-1.5cm}
\centering\includegraphics[width=.63\textwidth]{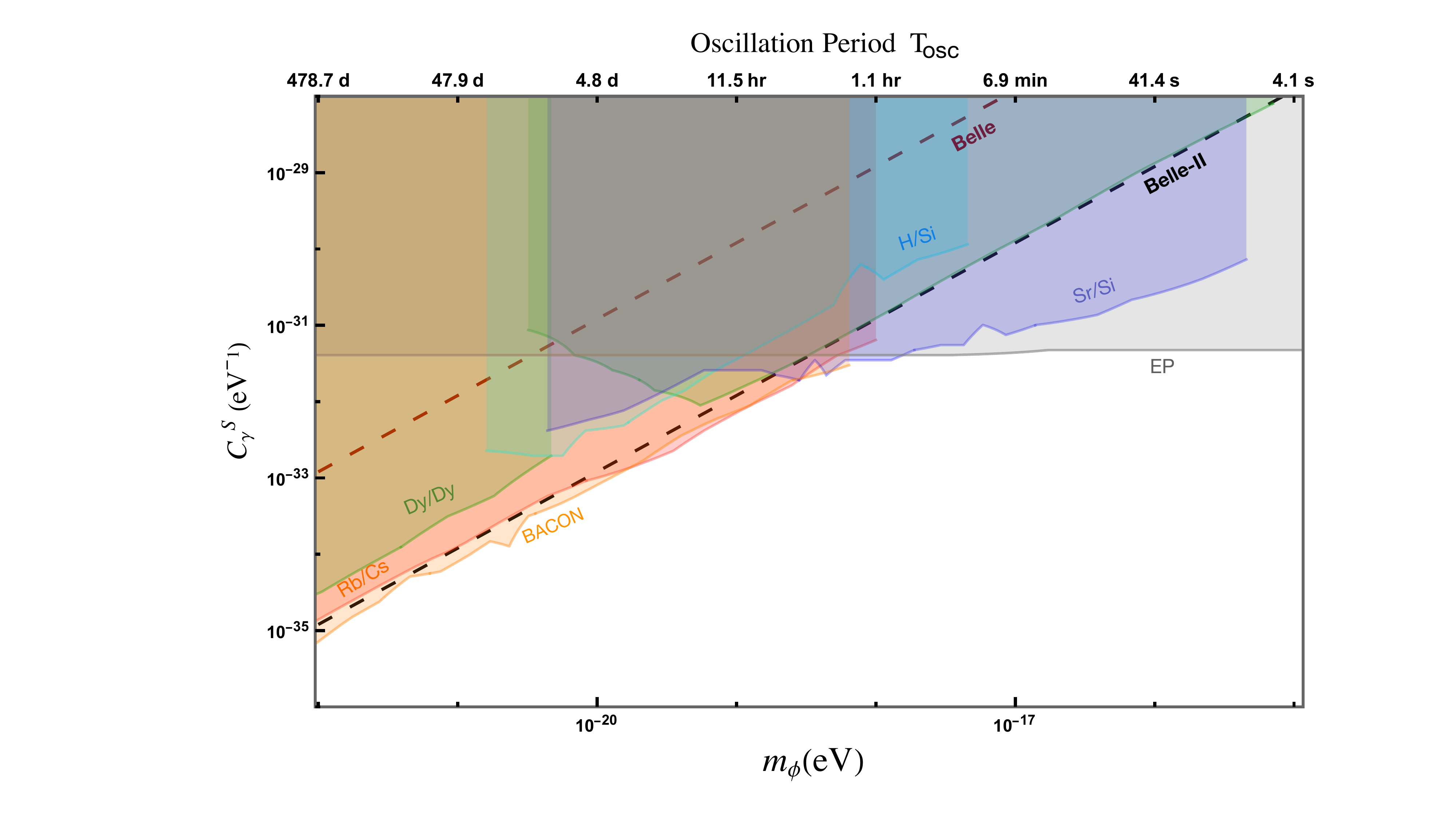}
       \hspace{-1cm}\raisebox{.15cm}{\includegraphics[width=0.61\textwidth]{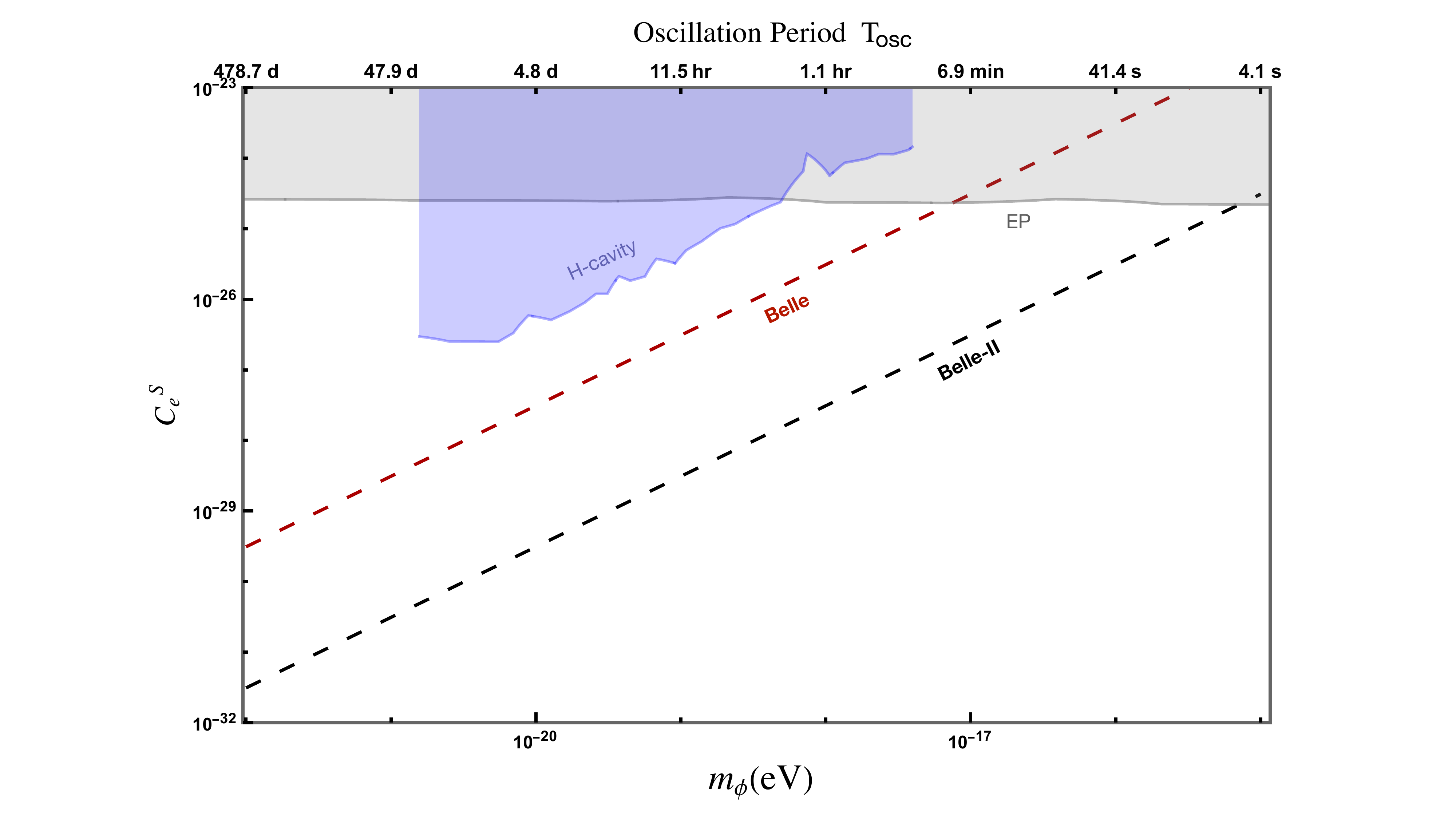}}
       \end{changemargin}
    \caption{The loop-induced photon and electron couplings for a scalar  $S$ in the ULDM background. The collider bounds correspond to Belle and Belle~II limits from $e^+ e^- \to S e^+ e^- \to\ell^+\ell^- e^+ e^-$ for 10\% mass modulation. The shaded lines denote the latest bounds from precision sensors such as atomic clocks (Rb/Cs, BACON), atomic spectroscopy (Dy/Dy), clock-cavity comparisons (H/Si, Sr/Si) and EP violation experiment (MICROSCOPE).}
    \label{fig:loopmodel3}
\end{figure}
In the limit $m_\phi\ll m_e$, we can use the low-energy theorem, so that
\begin{align}\label{eq:alphavar}
\mathcal{L}=\frac{\alpha}{8\pi}\sum_\psi b_\psi\left(\frac{\partial \ln m_\psi(\phi)}{\partial \phi}\Bigg\vert_{\phi=0}\right)\phi F_{\mu\nu}F^{\mu\nu}=\frac{\alpha}{6\pi}c_\psi \phi F_{\mu\nu}F^{\mu\nu}\,,
\end{align}
where we used the contribution to the beta function for a charged lepton, $b_\psi=4/3 N_c Q_\psi^2$. In the case of SM fermions charged under $U(1)_X$, the sum in \eqref{eq:alphavar} goes over all charged fermions, in the case of kinetic mixing instead, the sum extends over all charged SM fermions and should be extended to include the $W^\pm$ gauge bosons as well. Since we are going to focus on mediator masses below the electroweak scale, we can neglect these contributions.\footnote{For fermions heavier than the mediator the correct approach would be to integrate out the fermion box in the right panel of Fig.~\ref{fig:diagrams} first and matching to a Euler-Heisenberg type Lagrangian $\mathcal{L}= \displaystyle\frac{c_\text{EH}}{m_\psi^4}F_{\mu\nu}F^{\mu\nu}X_{\rho\sigma}X^{\rho\sigma}+\ldots$ before calculating the mediator loop.} 

The oscillation of the fine-structure constant induced by the mediator models considered here then reads
\begin{align}
\alpha^\text{eff}(\phi)=\alpha+\delta\alpha=\alpha(1-c_\gamma \phi)\,,
\end{align}
such that
\begin{align}
\frac{\delta \alpha}{\alpha} = \begin{cases}\displaystyle-\frac{1}{2\pi^3} \frac{\delta m_X}{m_X}\cos(\omega t)\,,& c_\psi=c_\psi^X,\\\displaystyle
-\frac{g_X^2}{12\pi^3} \frac{\delta m_X}{m_X}\cos(\omega t)\,,& c_\psi=c_\psi^\text{FS},\\\displaystyle
-\frac{\kappa_e^2}{2\pi^3} \frac{\delta m_S}{m_S}\frac{\mu_S^2}{m_S^2}\cos(\omega t)\,,& c_\psi=c_\psi^S.
\end{cases}
\end{align}
In Fig.~\ref{fig:loopmodel1} and Fig.~\ref{fig:loopmodel3}, we compare the limits from quantum sensors with bounds and projections from collider searches for the signature of the mediator resonance. The limits in Fig.~\ref{fig:loopmodel1} are for the model with a vector mediator with the Lagrangian $\mathcal{L}_1$ and a kinetic mixing term inducing interactions with the SM fermions. We fix the gauge coupling to $g_X=4\pi$ in this model, but the constraint from variations of $m_e$ or $\alpha$ can be obtained by rescaling the induced Wilson coefficients by $(g_X/4\pi)^2$. Note that $g_X$ does not enter these as a coupling constant here, but only determines the mediator mass $m_X=g_X v_D$. We compare constraints on couplings to photons (left panels) and electrons (right panels), respectively. In Fig.~\ref{fig:loopmodel3} we show the corresponding plot for the scalar mediator model. The limits are from frequency measurements with atomic clock and spectroscopy comparisons (Rb/Cs in red~\cite{Hees:2016gop,Antypas:2022asj}, the Boulder atomic clock optical network (BACON) in yellow~\cite{BACON:2020ubh} and Dy/Dy long-term and short-term measurements in green and light green~\cite{VanTilburg:2015oza}), clock-cavity searches (H/Si in cyan~\cite{Kennedy:2020bac}, Sr/Si in light purple~\cite{Kennedy:2020bac}) and also from the MICROSCOPE experiment testing the equivalence principal (EP in gray~\cite{Touboul:2022yrw}). In the case of the electron coupling, the light purple region is excluded from comparisons of a H-maser with a Si cavity~\cite{Kennedy:2020bac}, which isolates a functional dependence on $m_e$ in the measured cavity frequency ratio. 
In the case of a spin-1 mediator, we compare these limits to projections for a resonance search with $50\, {\rm ab}^{-1}$ for Belle-II based on $e^+ e^- \to \, \gamma X  \to  \gamma\ell^+\ell^- $~\cite{Ferber:2015jzj, Belle-II:2018jsg} and for $pp \to X \to \mu^+\mu^-$ at LHCb Run-6~\cite{Craik:2022riw}. For the scalar mediator, the collider bounds correspond to the search for a resonance with $626 \,{\rm fb}^{-1}$ of $e^+ e^- \to \, S  e^+ e^- \to  \ell^+\ell^- e^+ e^-$ at Belle~\cite{Cogollo:2024fmq} and the rescaled projection for Belle-II at $50\, \rm{ab}^{-1}$. 
The plot also shows the oscillation period as a function of dark matter mass. It is remarkable that collider searches are competitive and sometimes surpass the limits of the precision-driven constraints from atomic clocks. Note, that the effective couplings in Fig.~\ref{fig:loopmodel1} and ~\ref{fig:loopmodel3} are independent of the mediator mass $m_X$. In order to demonstrate the reach of colliders, we chose the value for $m_X$ for which the different searches provide the strongest constraint. 

In the context of the comparison to precision observables we want to emphasise the generality of the models discussed here. In \eqref{eq:LS} and \eqref{eq:gauge}, the interactions between the mediator and the SM fermions are scalar and vector-like interactions, respectively, and other coupling structures could be considered. However, neither the resonant production of mediator at a collider, nor the loop-induced corrections to electron mass or the fine-structure constant are affected significantly by different coupling structures. One could also consider models in which the dark matter field interacts only quadratically with SM fermions~\cite{Banerjee:2022sqg, Kim:2023pvt, Beadle:2023flm, Bauer:2024hfv}. In this case, the effective operators would be higher dimension and the corrections to SM parameters would be suppressed. However, the correction on the mediator mass \eqref{eq:LSL2L3delta1}-\eqref{eq:LSL2L3delta3} is also higher order in this case and therefore the relation between $\delta \alpha, \delta m_e$ and $\delta m_S$ or $\delta m_X$ remains parametrically the same as in the case. Further, quadratic interactions give rise to a screening effect that can suppress (or enhance) the field value close to a source such as earth~\cite{Hook:2017psm, Balkin:2020dsr, Bauer:2024yow}. In this case, additional constraints arise beyond those shown in Fig.~\ref{fig:loopmodel1} and Fig.~\ref{fig:loopmodel3}.

\section{Impact of dark matter oscillation on dark photon bounds at colliders}
\label{sec:collider}

In this section, we focus on the characteristics of dark matter oscillation and its impact on the searches at colliders and beam dumps. 
We are interested in ULDM masses for which the oscillation period spans a few hours to several months. 
This is a careful choice, since to have any impact on collider observables, the ULDM oscillation period ($T_{\rm osc}$) 
must be longer than the ``event time resolution'', characterised by the $\mathcal{O}({\rm ns})$``bunch crossing interval''
in colliders such as the LHC or Belle~II, or by the $\mathcal{O}(s)$``spill time'' in displaced-vertex searches such as SHiP. Each event thus probes a fixed ULDM phase, while the full dataset, collected over experimental integration times $T_{\rm int}$ of months to years, samples effectively random phases. Depending on whether $T_{\rm osc}$ is shorter, comparable to, or longer than $T_{\rm int}$, there can be different effects on the collider observables, which we will discuss later. It is, however, important to establish at this stage that the optimal ULDM mass range is
\begin{align}
    10^{-22} \lesssim m_\phi \lesssim 10^{-15}\ {\rm eV}\,,
\end{align}
corresponding to $T_{\rm osc}$ between 1.3~years to 4~seconds. 
The class of models we consider allows for time-dependent oscillations in the dark photon (or scalar) mass, as well as in the kinetic mixing parameter. Mass oscillations can leave observable imprints in collider-based resonant searches, manifesting as a smearing of the invariant mass distribution of final states such as dilepton pairs from dark photon decays. For displaced decay searches, however, mass oscillations do not affect the total signal yield. This is because the explicit mediator mass dependence cancels between the production rate and the decay probability within the detector shielding, leaving the expected number of observed events unchanged. In contrast, oscillations of the kinetic mixing parameter induce a residual, subleading modulation of the signal rate, which can impact displaced-vertex searches. We quantify this effect explicitly in the context of SHiP limits below.

In the following subsections, we therefore focus on two distinct scenarios with the example of a vector mediator interacting with the SM via kinetic mixing: (i) oscillating mediator mass and (ii) oscillating kinetic mixing.

\subsection{Oscillating mediator mass}
\label{sec:massmod}

To correlate modulation effects with collider limits, we define the maximum mass modulation ratio as:

\begin{align}
    \mathcal{R}_m = \frac{m_X(t)-m_X^0}{m_X^0}\propto\frac{1}{\Lambda}\frac{\sqrt{2\rho_{\rm DM}}}{m_\phi}\cos\omega t
    \label{eq:uldm}
\end{align}

As a reasonable estimate, we take $\mathcal{R}_m \sim 10\%$, which from the above expression explicitly depends on the ULDM mass ($m_\phi$) and the coupling strengths ($1/\Lambda$) respectively. For a GeV-scale mediator mass, this corresponds to coherent oscillations with an amplitude ranging from several hundred MeV to a few GeV, oscillating around the central mass at the Compton frequency of $\phi$. The typical search strategies for dark photons in this mass range include bump-hunt analyses in high-luminosity collider experiments such as LHCb and Belle~II.

Each signal event at these experiments corresponds to an individual $e^+ e^-$ or $p\,p$ collision producing the mediator, followed by a prompt decay into a pair of electrons or muons. The events are recorded over a timescale of a few nanoseconds, set by the bunch crossing interval. At Belle-II, this interval is approximately $\sim$ 4--8 ns, depending on the specific run configuration~\cite{Martini:2018hhh} and slightly larger in LHCb, ie, $\sim 25\ {\rm ns}$~\cite{Franzoso:2021yjn}. Therefore, for our choice of ULDM mass range, on the timescale of an individual event, the phase of oscillation can be treated as constant over a single ULDM oscillation cycle. However, across the full data-taking period spanning several years, colliders sample the oscillating mediator mass at effectively random phases. Each recorded event acts as a ``snapshot" with the DM background \emph{frozen} at fixed instants randomly distributed over the oscillation cycle. 

This implies that colliders do not directly \emph{see} the time variation of the mediator mass or decay width; rather, they record events occurring at random phases of the ULDM oscillation. Two distinct phenomenological consequences arise. First, when the central mediator mass lies well above the kinematic threshold for dilepton decays, a large modulation amplitude spreads the resonance across a wide invariant-mass window, flattening the distribution and suppressing the apparent peak height. Second, when the central mass lies near or just below the decay threshold, the modulation can periodically push $m_X(t)$ above threshold during parts of the oscillation cycle, temporarily opening the decay channel and producing detectable events even when the static mass would be kinematically forbidden.
 We quantify the two effects below for prompt dimuon decays in LHCb (Run-2), followed by dilepton decays in future colliders such as LHCb (Run-6) and Belle-II.

\subsubsection{Modulation effects above the kinematic threshold}

Traditional bump-hunt searches model the invariant mass ($m_{\ell\ell}$) peak for a resonance decaying promptly into $\ell^+\ell^-$ pairs as a Gaussian-smoothed delta function, with the width set by detector resolution. Belle-II typically has a mass resolution of $\Delta m_{\rm res} \sim 2.5-5\ {\rm MeV}$~\cite{Belle-II:2024wtd} for dark photons decaying into dimuons, whereas for LHCb, it varies between $0.7~{\rm MeV} - 0.7~{\rm GeV}$ from the dimuon threshold to 70 GeV dark photon mass~\cite{lhcb}. Since collider limits on resonances are exclusion limits arising from the non-observation of a resonance peak in the invariant mass spectrum, a modulation amplitude larger than the mass resolution can still be compatible with current analysis strategies.

If the mediator mass modulation exceeds the detector’s mass resolution $\delta m_X > \Delta m_{\rm res}$,  the resonance peak in the invariant mass spectrum is broadened. As a result, the signal is no longer concentrated within a single bin but is distributed over multiple mass bins, even if the total width of the mediator is narrow in the absence of the dark matter background field. Due to the oscillatory nature of the modulation, the resonance spends only a fraction of its cycle within any given bin. Consequently, the event rate in each bin is effectively reduced. Since the modulation phase is unknown at the time of measurement, one must average over the full oscillation cycle to estimate the total number of signal events accurately.

The total number of signal events depends on the dimuon production cross-section $\sigma$, integrated luminosity $\mathcal{L}$, the detector efficiency $\eta_{\rm eff}$, and the decay branching fraction for its decay into detectable final states. For the dimuon channel, the relationship can be expressed as
\begin{align}
    N_{\rm obs}\, \propto\, \sigma(pp/e^+e^-\to X \to \mu^+\mu^-)\times\eta_{\rm eff}\times\mathcal{L}\,,
    \label{eq:Nobs}
\end{align}
where the dominant production mechanism for dark photons decaying into dimuons is Drell-Yan~\cite{Graham:2021ggy} in LHCb and Brehmstrahlung or initial state radiation~\cite{Belle-II:2018jsg} in Belle-II. The decay width of a kinetically mixed vector mediator into leptons scales as $\Gamma_X\, \propto\, \epsilon^2m_X$~\cite{Chakraborti:2021hfm} and in the narrow-width approximation ($\Gamma_X \ll m_X$), the cross-section takes the form
\begin{align}
   \sigma\, \propto\, \frac{\epsilon^4}{(m_X^2-q^2)^2+m_X^2\Gamma_X^2}\times {\rm BR(X\to \mu^+\mu^-)}\approx \frac{\epsilon^4\,\delta(m_X^2-q^2)}{m_X \Gamma_X}\times {\rm BR(X\to \mu^+\mu^-)}\nonumber\\\approx \frac{\epsilon^2}{m_X^2}\,\delta(m_X^2-q^2)\times {\rm BR(X\to \mu^+\mu^-)}\,,
\end{align}
where $q$ is partonic or leptonic momenta, depending on the type of collider \footnote{At LHCb, we assume the partonic-level cross-section suffices, since the hadronic cross-section with PDF convolution preserves the essential $1/m_X^2$ scaling behavior.}. For mediator masses well above the dimuon threshold and below the opening of additional decay channels, the branching ratio ${\rm BR}(\mu^+\mu^-)$ is approximately constant. As a result, mass oscillations do not significantly affect the branching fraction unless a kinematic threshold is crossed. Even in that case, provided $\delta m_X \ll m_X$, threshold effects are averaged over the oscillation cycle and remain subleading compared to the dominant modulation of the production rate. Consequently, when taking the ratio of the modulated to static signal rates to reinterpret limits on $\epsilon$, the decay branching fraction cancels to leading order, and the impact of mass modulation is governed primarily by the production rate scaling.

In the ULDM background, due to time-dependent mass modulation, the resonance peak in the invariant mass distribution shifts around the detector-sensitive region during the oscillation cycle. Since the mediator mass is unknown, a time-averaged signal estimate becomes necessary. To model this, we discretise the oscillation phase as $\theta=\omega t$ and compute the phase-averaged signal using
\begin{align}
    \langle S_{\rm obs}\rangle\, \propto\, \frac{1}{2\pi}\int_0^{2\pi}\sigma(m_X(\theta))\,\Theta\left(|m_X(\theta)-m_X^0|-\Delta m_{\rm res}\right)d\theta 
    \label{eq:sobs}
\end{align}
where $\Theta$ is a step-function ensuring that only those intervals where the modulated mass lies within the resolution window contribute to the observed rate. 
Upon taking the ratio of the modulated signal rate to the standard no-modulation or ``static" scenario, we can quantify the effect on the sensitivity to the dark photon coupling. Since the phase-averaged signal yield is always smaller than the static case in each bin, a larger coupling is required to produce the same number of observable events. This leads to the following relation
\begin{align}
  \epsilon_{\rm osc}=\sqrt{  \frac{S_{\rm static}}{\langle S_{\rm obs}\rangle}}\times \epsilon_{\rm static}
  \label{eq:eps1}
\end{align}
where $S_{\rm static}$ denotes the expected signal yield in the absence of mass modulation, which corresponds to a Gaussian resonance only smeared by detector resolution, as in the standard bump-hunts. The parameters $\epsilon_{\rm static}$ and $\epsilon_{\rm osc}$ represent the bounds on the kinetic mixing parameter in the static and modulated scenarios, respectively.
\begin{figure}[ht]
    \centering
    \includegraphics[width=0.75\linewidth]{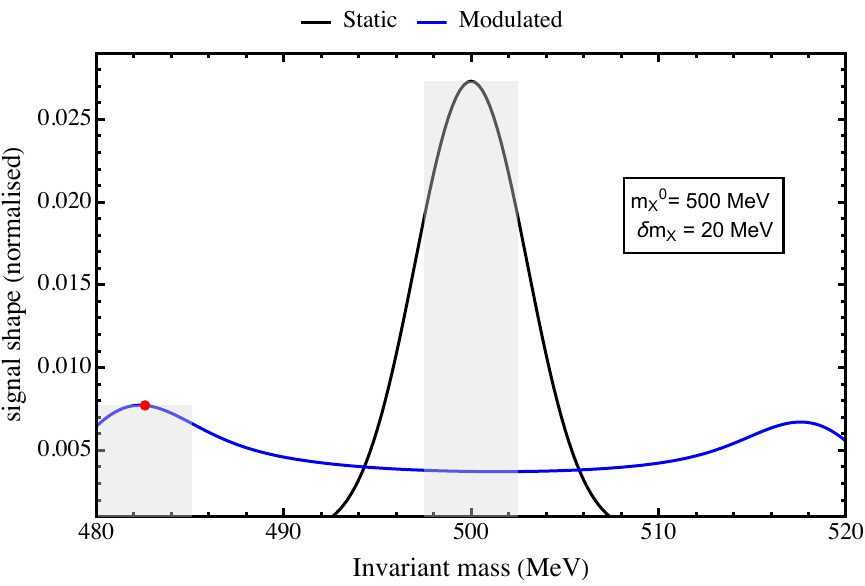}
    \caption{Normalised invariant mass distributions for a static dark photon (black) and in the presence of an oscillating DM background (blue) for a central mass $m_X^0=500\ {\rm MeV}$ and modulation amplitude $\delta m_X=20\ {\rm MeV}$. The gray band indicates the detector resolution around the central mass. In the oscillating case, the resonance is spread across the modulation window, diluting the signal within any single mass bin. The red marker highlights the tallest bin (local maximum) in the modulated distribution, which is located near the edge of the modulation window rather than at the central mass.}
    \label{fig:comphisto}
\end{figure}
We develop a simple Monte-Carlo based approach to quantify the modulation. In Fig.~\ref{fig:comphisto}, we compare the invariant mass distributions for the static (standard) case in black and the modulated case in blue. The central dark photon mass is fixed at $m_X^0=500\ {\rm MeV}$ with the modulation amplitude, $\delta m_X =20\ {\rm MeV}$. The gray-shaded region for both cases corresponds to the detector-resolution convolution, where we use a fixed $\Delta m_{\rm res}= 2.5\ {\rm MeV}$. As expected, the phase-averaged signal, averaged over a full oscillation cycle, is spread over the oscillation window of $2\delta m_X = 40\ {\rm MeV}$. The double peaks at $m_X^0\pm \delta m_X$ are a classic feature of the mass modulation as the oscillating field spends more time near the edges of the modulation window, compared to the center. For oscillations of the form $m_X(\theta)=m_X^0+\delta m_X\cos\theta$, the distribution of signal counts follows the probability density function 
\begin{align}
P(m_X) = \left| \frac{d\theta}{dm_X} \right| = \frac{1}{\sqrt{\delta m_X^2 - (m_X - m_X^0)^2}}\,,
\label{eq:Pmod}
\end{align}
which has a minimum at $\cos\theta=0$ and diverges for $\cos\theta=\pm1$, although, in a realistic histogram with finite resolution, the divergence is smoothed out, and the two extrema appear as broadened peaks rather than true singularities. While from ~\eqref{eq:Pmod} one expects the two peaks at the modulation extrema to be identical in height, the actual phase-averaged signal is also weighted by $1/m_X(\theta)^2$ due to the $\sigma(m_X(\theta))$ factor in ~\eqref{eq:sobs}. This asymmetric weighting renders the peak at $m_X^0-\delta m_X$ taller than the one at $m_X^0+\delta m_X$, consistent with the red marker in the figure, which identifies the tallest bin in the invariant mass distribution.

This is a key feature of our analysis, as it implies that in standard bump-hunt techniques, where a mass range is scanned for resonant peaks, the tallest peak in the invariant mass distribution may appear shifted from the expected central mass if an oscillating DM is present in the background, leading to an oscillating resonance in the $m_{\ell\ell}$ distribution. The underlying ULDM background parameters directly determine the exact shift. Moreover, a larger modulation window would result in a broader spread, causing fewer events per bin, and suppressing the modulated peak height further relative to the static case. For sufficiently large $\delta m_X$, the peak may remain entirely undetected due to an inadequate signal-to-noise ratio. 
\vspace{1em}
\begin{figure}[htb!]
    \centering
    \includegraphics[width=0.75\linewidth]{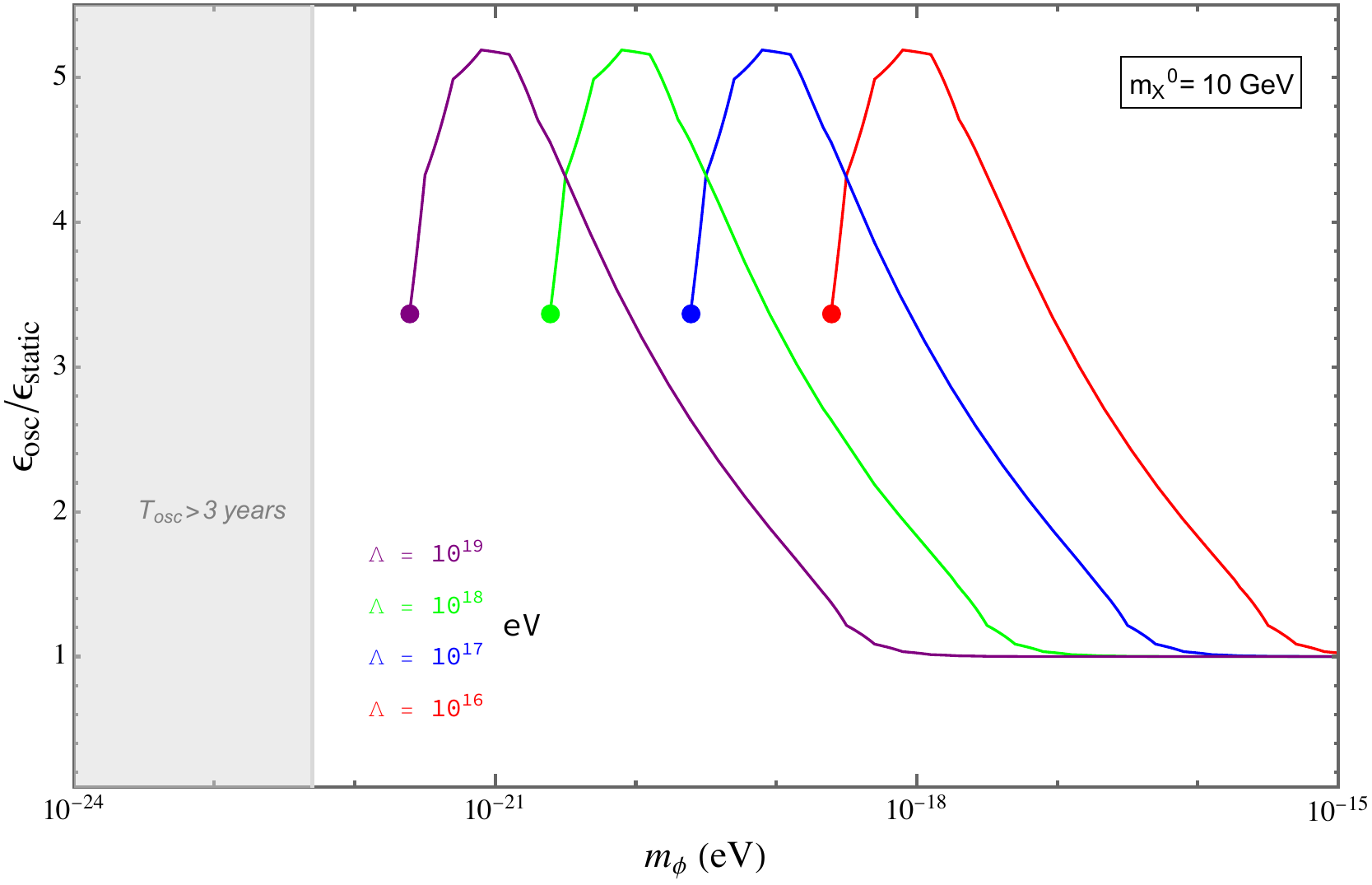}
    \caption{Ratio of the kinetic mixing parameter in the oscillating case to the static case $\epsilon_{\rm osc}/\epsilon_{\rm static}$ for a 10~GeV dark photon at Belle~II, shown as a function of the ULDM mass $m_\phi$. Different coloured curves correspond to different values of the ULDM energy scale $\Lambda$, quantifying the DM coupling strength with the SM. The grey shaded region denotes oscillation periods longer than the experimental run-time ($T_{\rm osc} \gtrsim\ 3\ {\rm years}$, where the modulation is effectively static). The coloured markers indicate the ULDM masses at which the modulation is 50\% for each $\Lambda$. At larger $m_\phi$, the oscillation period becomes short compared to the data-taking time, and the effect of modulation averages out, causing the ratio to approach unity.}
    \label{fig:uldm}
\end{figure}

To quantify the oscillating resonance impacting collider sensitivities on the dark photon kinetic mixing parameter, we show in Fig.~\ref{fig:uldm} how the ratio $\left(\epsilon_{\rm osc}/\epsilon_{\rm static}\right)$ depends on ULDM mass and coupling strength via the mass modulation amplitude. The central/static dark photon mass is fixed at $m_X^0=10\ {\rm GeV}$ and each colour corresponds to a specific DM coupling strength denoted by $d_H/\Lambda$. Decreasing $m_\phi$ gradually for a fixed $\Lambda$ translates to an increasing $\delta m_X$ (or $\mathcal{R}_m$), as is straightforward from ~\eqref{eq:LSL2L3delta1}. This explains why, from right to left, each coloured line begins with a plateau which corresponds to $\epsilon_{\rm osc}/\epsilon_{\rm static}=1$ and then rises as the modulation becomes noticeable to cause any significant change in the signal distribution. However, the ratio reaches a maximum for an optimum $\mathcal{R}_m$ and then begins to fall as the modulation ratio grows further. This is due to two competing effects: (i)~For a fixed central mass, increasing $\delta m_X$ leads to dilution of events across a broader modulation window, therefore reducing the peak height. This leads to relaxing the bound with an increase in $\epsilon_{\rm osc}/\epsilon_{\rm static}$. (ii)~An increasing $\delta m_X$ also leads to a smaller $m_X(\theta)$  in the lower edge of the modulation window, which boosts up the height of the resonance peak due to the $1/m_X(\theta)^2$ weighting in the total number of signal events. For large enough $\mathcal{R}_m$, the second effect overpowers the first, thereby reducing the contrast between the static and the modulated peak height. This results in a reduction in $\epsilon_{\rm osc}/\epsilon_{\rm static}$. Since the transition between two effects depends solely on $\delta m_X$ (or $\mathcal{R}_m$) and not on the ULDM parameters, the optimum $\mathcal{R}_m$ remains constant for all choices of $d_H/\Lambda$ and we estimate it to be $\mathcal{R}_m^{\rm opt}\approx 12\%$ for our choice of parameters. 

There is, however, a shift along the $m_\phi$ axis as a function of $\Lambda$. A small $\Lambda$ (or large $d_H/\Lambda$) requires a large $m_\phi$ to keep the $\delta m_X$ fixed, as is obvious from ~\eqref{eq:LSL2L3delta1}. This is consistent with why the stretch of the plateau is the smallest for the red line and why the peak corresponding to $\mathcal{R}_m^{\rm opt}$ occurs at a $m_\phi$ value larger than the blue, green and purple lines, which all correspond to smaller $d_H/\Lambda$ values than the red line. Finally, we stop the computation when $\mathcal{R}_m \approx 50 \%$ for each line, beyond which the modulation is too strong for our signal extraction prescription to remain valid. This is denoted by the coloured dots for each line.

\subsubsection{Mass modulation effects around the kinematic threshold}

For a central mediator mass around the dimuon threshold, the optimal modulation  implies $\delta m_X\approx 10-20\ {\rm MeV}$, which is comparable to the detector resolution in colliders. As discussed before, $\delta m_X \lesssim \Delta m_{\rm res}$ does not typically lead to signal weakening since the modulation is within the detector resolution, effectively rendering the signal static. However, in the vicinity of kinematic thresholds, even a small modulation can produce a significant impact by periodically opening and closing decay channels over an oscillation cycle. In the absence of time-resolved data, this behaviour does not manifest as periodic bursts of events in the detector, but instead results in a net reduction of the time-averaged number of decays falling within the detector acceptance and mass resolution window. 

In particular, when the central mass $m_X^0$ lies close to the threshold, the instantaneous mass $m_X(\theta)$ can cross the kinematic boundary during part of the oscillation cycle. As a result, events contribute intermittently to regions of phase space that are either fully allowed or fully forbidden in the constant mass case. This leads to a partial population of the near-threshold phase space and effectively smooths the transition between the excluded and allowed regions assumed in the static analyses. Consequently, experimental limits derived under the assumption of a fixed mediator mass may be quantitatively modified when mass modulation is present.

We show this near-threshold effect by reinterpreting the current LHCb prompt dark photon search~\cite{Graham:2021ggy}, which targets dimuon final states exclusively for dark photon masses above the dimuon threshold (i.e. $m_{\rm th}\simeq 211.32\ {\rm MeV}$), up to $\sim$~70 GeV. In the presence of an oscillating DM background, however, even when $m_X^0$ is below the dimuon threshold, the instantaneous mediator mass $m_X(t)$ can periodically exceed it. During these intervals, the decay $X\to\mu^+\mu^-$ becomes kinematically allowed, leading to a nonzero contribution to the observable signal rate. 

For example, a $10\%$ modulation around $m_{\rm th}$ corresponds to a modulation window of $2\delta m_X \approx 42\ {\rm MeV}$, such that dimuon decays are intermittently permitted for central masses as low as $m_X^0\gtrsim m_{\rm th}-\delta m_X$. As a result, experimental sensitivity extends partially below the nominal threshold assumed in the standard (static) analyses.

Nevertheless, consistency with the absence of observed dimuon events below threshold requires the effective kinetic mixing $\epsilon_{\rm osc}$ in this regime to be sufficiently small, ensuring that the time-averaged signal yield remains below the experimental detection threshold. Consequently, while mass modulation can relax the strict kinematic cut-off of static searches, it does not generically imply strong exclusions below $m_{\rm th}$, but rather motivates a modified interpretation of near-threshold limits.

To quantify the rescaling of limits, we introduce below the quantity $f_{\rm ON}$, which is the fraction of time the decay is kinematically allowed over a period of oscillation(see Section~\ref{sec:frac} for the derivation)
\begin{align}
    f_{\rm ON}=\frac{1}{\pi}\cos^{-1}\left(\frac{m_{\rm th}-m_X^0}{\delta m_X}\right)\,.
\end{align}
The number of events in ~\eqref{eq:Nobs} must then be scaled with the above fraction as:
\begin{align}
    N_{\rm obs}\, \propto\, \epsilon_{\rm osc}^2\times  \eta_{\rm eff}\times\mathcal{L}\times f_{\rm ON}\,.
    \label{eq:Nobs1}
\end{align}
By comparing $N_{\rm obs}$, to a reference value $N_{\rm th}$, (\emph{e.g.} the expected yield at threshold), we obtain an estimate of  $\epsilon_{\rm osc}$ relative to the $\epsilon$ right at the threshold in static analyses. A conservative bound $N_{\rm obs} \lesssim 3$ is used to represent the non-observation of events.\\
Therefore, the rescaling of the LHCb exclusion limits in the oscillating case can be written as
\begin{align}
    &f_{\rm ON}\times\left(\frac{\epsilon_{\rm osc}}{\epsilon_{\rm th}}\right)^2 = \frac{N_{\rm obs} }{N_{\rm th} }\,,\nonumber\\
   &\epsilon_{\rm osc} \lesssim \sqrt{\frac{3}{f_{\rm ON}\times N_{\rm th}}}\times\epsilon_{\rm th}\,.
\end{align}
It is also worth noting that this threshold effect also affects the mass window $m_{\rm th}\lesssim m_X(t)\lesssim m_{\rm th}+\delta m_X$ where the number of observed events with DM background modulation is significantly less than in the static case, because $X\to \mu^+\mu^-$ is absent for parts of the oscillation cycle. To match the expected signal yield from the static case, $\epsilon_{\rm osc}$ must be relaxed. For this range we used the following scaling
\begin{align}
    \frac{N_{\rm obs}}{N_{\rm static}}&\approx f_{\rm ON}\times\left(\frac{\epsilon_{\rm osc}}{\epsilon_{\rm static}}\right)^2=1\,, \nonumber\\
    \epsilon_{\rm obs}&\simeq \frac{\epsilon_{\rm static}}{\sqrt{f_{\rm ON}}}\,.
\end{align}

\begin{figure}[tb!]
    \centering
    \includegraphics[width=0.85\linewidth]{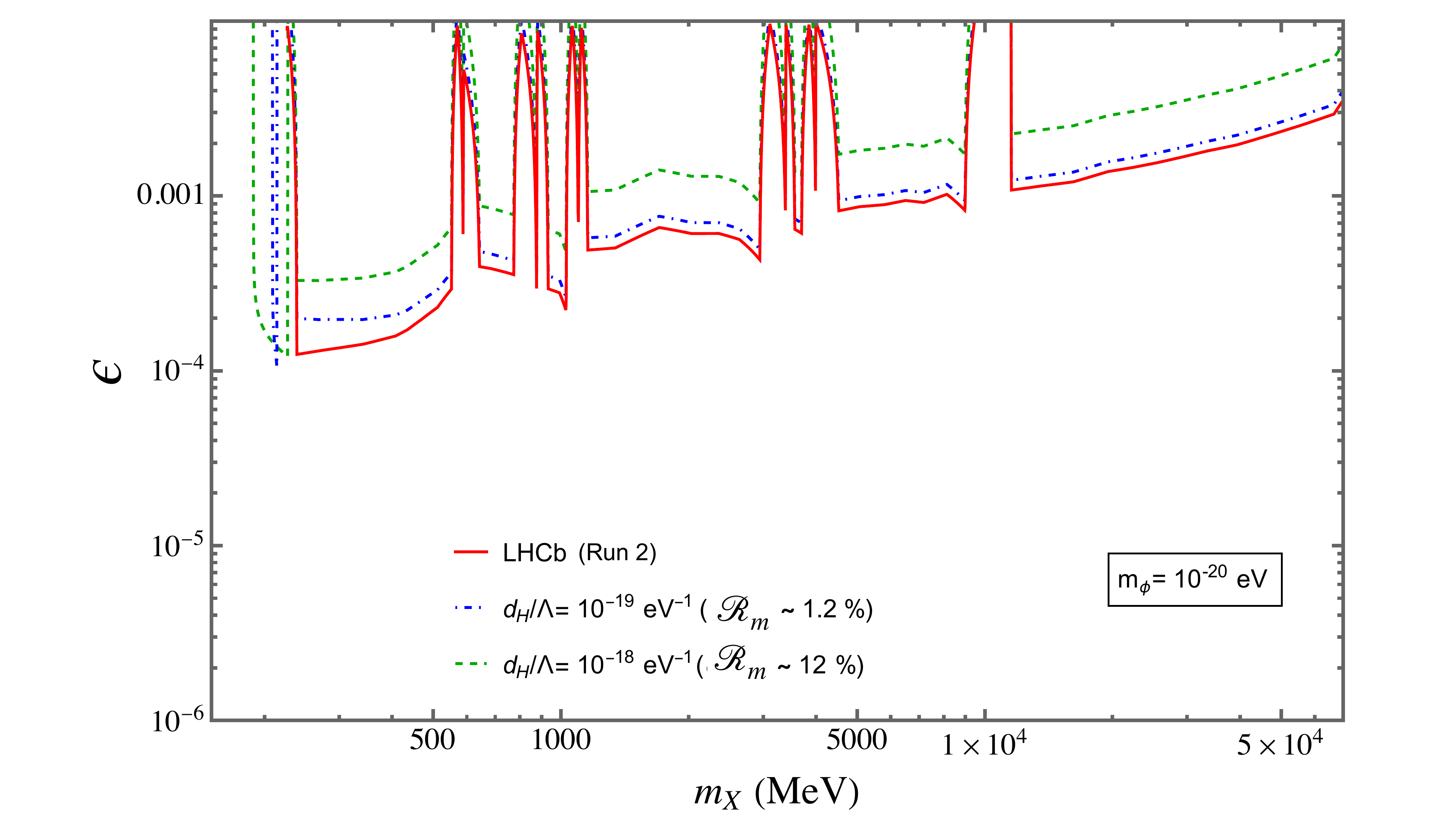}
    \caption{Oscillating resonance effect on current LHCb prompt-search bound on dark photon parameter space. The red solid line corresponds to the current limits, whereas the weakened bounds due to modulation are shown in green and blue, each corresponding to different ULDM couplings, leading to different percentages of the mass modulation.}
    \label{fig:lhcb}
\end{figure}

Fig.~\ref{fig:lhcb} summarises the impact of oscillating DM backgrounds on existing LHCb prompt dimuon limits~\cite{LHCb:2019vmc,Graham:2021ggy} across the full dark photon mass range, combining the effects discussed above. For mediator masses well above the dimuon threshold, the dominant effect arises from mass modulation exceeding the detector resolution, leading to a dilution of the resonance across multiple invariant mass bins and reduction of the per-bin signal significance. For the LHCb analysis, we model the mass resolution by linearly interpolating between the quoted values of  $0.7~{\rm MeV}$ near the dimuon threshold and $0.7~{\rm GeV}$, at a dark photon mass of 70~GeV~\cite{lhcb}. We show the weakening of the exclusion limits for two representative modulation ratios $\mathcal{R}_m=1.2\%$ (blue dashed) and $\mathcal{R}_m=12\%$ (green dashed), respectively, while the static limit is shown by the solid red line. The green curve corresponds to the optimal modulation regime identified by the peaks in Fig.~\ref{fig:uldm}, where the resonance dilution is maximal, and the resulting weakening of $\epsilon$ reaches nearly a factor of 5. We also incorporate the near-threshold effect described above, whereby mass modulation allows the exclusion contours to extend to slightly lower mediator masses, determined by the modulation amplitude $\delta m$, compared to the static case.

\vspace{1cm}

\subsection{Oscillating kinetic mixing parameter}
\label{sec:epsmod}

In addition to prompt resonance searches, oscillating DM backgrounds can also impact displaced vertex searches in beamdump experiments such as SHiP~\cite{SHiP:2025ows} and E137~\cite{Batell:2014mga}. However, unlike prompt searches, the total signal yield in displaced decays is largely insensitive to the mediator mass once the decay length is macroscopic. As a result, mass oscillations do not significantly modify displaced vertex sensitivities at leading order. Instead, the dominant effect arises from time-dependent variations of the kinetic mixing parameter, as realised in models described by ~\eqref{eq:dim5eps} and \eqref{eq:osc_eps}. In this subsection we therefore focus on the oscillating $\epsilon$ scenario as the suitable testbed for probing ULDM effects in displaced vertex searches.\\
In a SHiP-like displaced decay search setup, the number of signal events is given by~\cite{Bauer:2018onh}
\begin{align}
    N\, \propto \int dE_{X}\, \sigma(E_X)\, P_{\rm{dec}}(E_X) ,
    \label{eq:Nship}
\end{align}
where $P_{\rm{dec}}$ is the probability that a dark photon $X$ decays within the detector volume 
\begin{align}
    P_{\rm{dec}}(E_X) = 
    e^{-L_{\rm{sh}} / \ell_X }
    \left( 1 - e^{-L_{\rm{dec}} / \ell_X} \right),
\end{align}
with $L_{\rm sh}$ and $L_{\rm dec}$ signifying the lengths of the shielding and the detector, respectively, and the corresponding decay length in the lab-frame is given by
\begin{align}
    \ell_X = \gamma_X c \tau_X 
    = \frac{E_X}{ m_X}\, \frac{1}{\Gamma_X} .
\end{align}
where as discussed before, the decay width scales as $\Gamma_X\, \propto\, \epsilon^2 m_X$. \\
The production cross-section for a beam of protons on a thick target can be approximated as
\begin{align}
    \sigma\, \propto\, 
    \frac{\alpha^3\, \epsilon^2\, Z^2}{m_X^2}\,,
\end{align}
 where $Z$ is the atomic number of the target nucleus and $\alpha$ denotes the fine-structure constant.

 Upon substituting $P_{\rm dec}$ ($\simeq L_{\rm dec}/\ell_X$ for the standard assumption $\ell_X \gg L_{\rm sh},  L_{\rm det}$ ) and $\sigma$ in ~\eqref{eq:Nship}, we find that the $m_X$ dependence explicitly drops out at leading order from the total number of events. Therefore, the mass oscillation scenarios are ineffective to impact these searches. On the other hand, due to the strong $\epsilon$ dependence in the total yield ($N \propto \epsilon^4$), any time-modulation of the kinetic mixing parameter could directly lead to observable variations in the event rate.
 
SHiP operates with the CERN SPS accelerator beam, which can deliver an annual yield of up to $4\times10^{19}$ protons on target in slow-extraction mode, with one-second-long spills of de-bunched beam at a momentum of $400~\rm{GeV}/c$~\cite{SHiP:2021nfo,SHiP:2025ows}. For signal observation, this corresponds to a recording window of comparable duration, during which the decays occurring outside the shielding are detected. 

In the SHiP sensitivity region for dark photons, $10^{-8} \lesssim \epsilon \lesssim 10^{-3}$, the corresponding proper lifetime spans a wide range, typically $ 10^{-16}\ \rm{s}\lesssim \tau_X \lesssim 10^{-6}\ \rm{s}$ (corresponding to the benchmark value $m_X^0=500\ {\rm MeV}$), but remains orders of magnitude shorter than the spill duration. As a result, individual dark photon decays are effectively instantaneous on experimental timescales, and no amount of decay width or lifetime modulation can cause a single decay to extend over a spill time.

The production cross section, on the other hand, gets a sub-leading effect from oscillating $\epsilon$ due to its scaling as $\sigma \propto \epsilon^2$, implying that the number of dark photons produced during each spill oscillates in time with $\epsilon(t)$. For our choice of $m_\phi$, the ULDM oscillation period always satisfies $T_{\rm osc} \gg T_{\rm spill}$. Therefore, each spill effectively samples a random phase of $\epsilon(\theta)$. However, because SHiP operates with a large number of spills per day ($\sim 5000$) and collects data over several years, the cumulative phase coverage becomes effectively uniform. As a result, the modulation averages out in the total number of observed events.

To quantify the oscillation effects, we express the time-averaged number of events in the presence of an oscillating kinetic mixing $\epsilon(t)$ as
\begin{align}
    \langle S_{{\rm obs}} \rangle 
    = \frac{1}{T_{\rm osc}} 
    \int_0^{T_{\rm osc}} 
    dt_0 \int_{t_0 - T_{{\rm spill}}/2}^{t_0 + T_{{\rm spill}}/2} 
    dt\, 
    \sigma\!\big(\epsilon(t)\big)\, 
    P_{\rm dec}\!\big(\epsilon(t)\big),
\end{align}
where $t_0$ labels the central phase of the spill, and the inner integral accounts for the finite duration of each spill.

As discussed above, the modulation does not affect the decay probability for values of $\epsilon$ typically within the SHiP sensitivity range. However, for the central value of the kinetic mixing parameter $\epsilon^0$ just outside this range, the slow oscillation of the decay length $\ell_X(t)\, \propto\, 1/\epsilon^2(t)$ can have a threshold effect on the sensitivity. During phases of the oscillation cycle, $\ell_X$ may periodically fall within the detector acceptance, allowing a fraction of dark photons to decay between the shielding and the detector volume. 

Ensuring that only those produced dark photons satisfying the geometric and kinematic acceptance contribute to the signal can be implemented using a  $\Theta$-function that enforces the visibility condition,
\begin{align}
\Theta_{\rm vis}\big(t\big) = 
\Theta\!\big(\ell_X(t) - L_{\rm sh}\big)\,
\Theta\!\big(L_{\rm det} - \ell_X(t)\big),
\end{align}
where $\ell_X(t)$ denotes the phase-dependent decay length. 

Including this acceptance factor, the time-averaged signal yield becomes
\begin{align}
    \langle S_{\rm obs} \rangle 
    = \frac{1}{T_{\rm osc}} 
    \int_0^{T_{\rm osc}} 
    dt_0
    \int_{t_0 - T_{\rm spill}/2}^{t_0 + T_{\rm spill}/2} 
    dt\, 
    \sigma\!\big(\epsilon(t)\big)\,
    \Theta_{\text{vis}}\!\big(t\big) 
    P_{\text{dec}}\!\big(\epsilon(t)\big),
    \label{eq:Nmod_theta}
\end{align}
where the inner integral represents the convolution over the spill duration in the production rate, and $\Theta_{\rm vis}$ incorporates the phase-dependent geometric acceptance.

\begin{figure}
    \centering
    \includegraphics[width=0.75\linewidth]{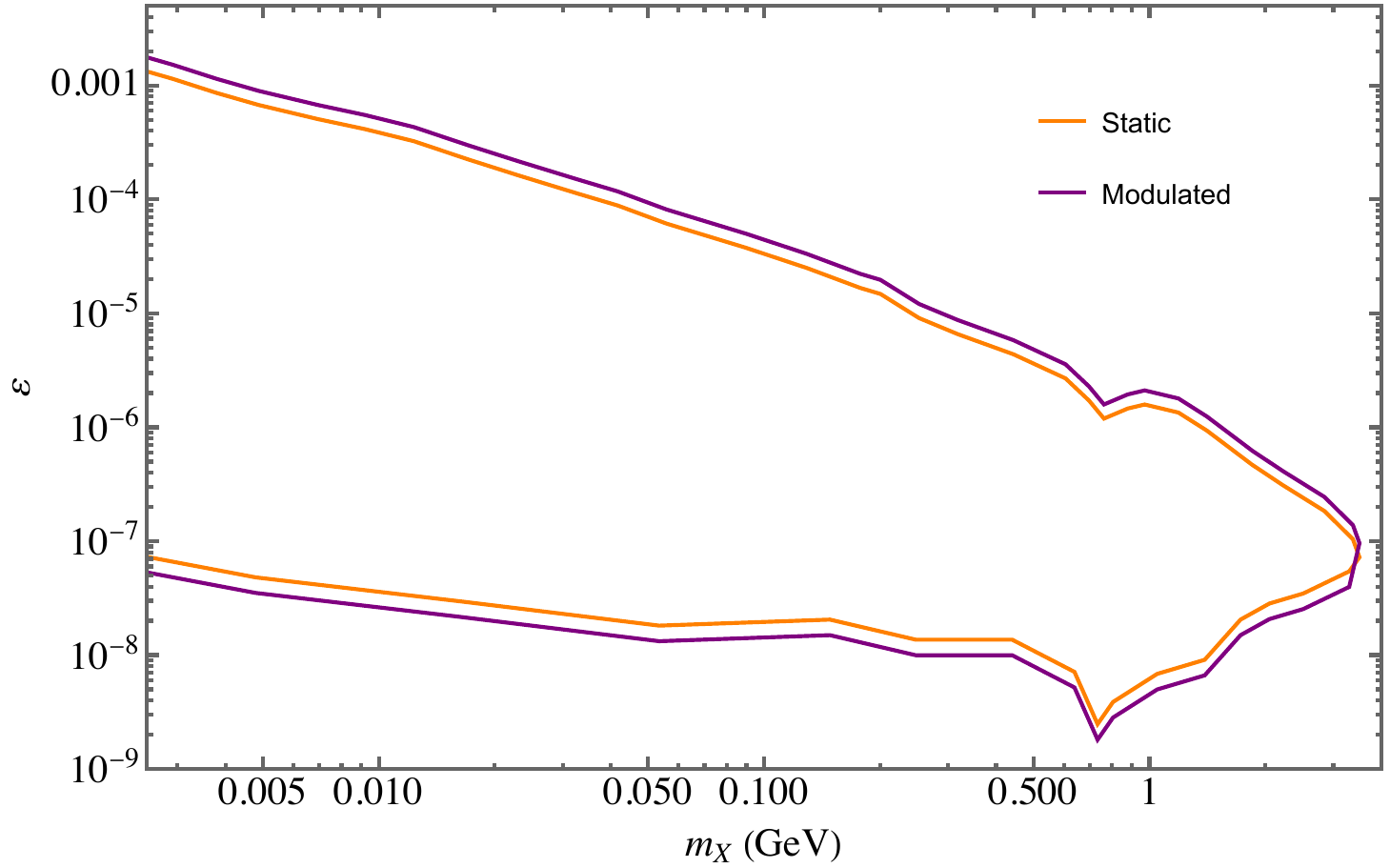}
    \caption{SHiP limits on the kinetic mixing parameter $\epsilon$ for static (orange) and oscillating (purple, 50\% modulation) scenarios. The modulation affects regions where the detector acceptance is controlled by decay length thresholds, leading to an increased sensitivity in the short-lived regime (larger $\epsilon$) as well as increased sensitivity in the long-lived regime (smaller $\epsilon$). Away from these acceptance edges, the impact of modulation is negligible, and the limits track the static case.}
    \label{fig:ship}
\end{figure}

 We implement ~\eqref{eq:Nmod_theta} to assess how an oscillating $\epsilon(t)$ affects the sensitivity of SHiP-like displaced-vertex searches. In Fig.~\ref{fig:ship} we illustrate the maximal effect, taking a $50\%$ modulation in $\epsilon$. As discussed above, away from the exclusion boundaries, i.e., where the central value of the kinetic mixing paramter $\epsilon^0$ lies well inside the SHiP sensitivity band, the time average essentially washes-out the modulation because the yield scales as $N\,\propto\,\epsilon^4$ and the integrand in ~\eqref{eq:Nmod_theta} varies smoothly over a spill and over many ULDM cycles. In contrast, near the edges the acceptance is controlled by the decay-length window enforced by the step factors in ~\eqref{eq:Nmod_theta}, $L_{\rm sh}<\ell_X(t)<L_{\rm sh}+L_{\rm det}$ with $\ell_X\,\propto \,1/\epsilon^2$. A sizable modulation therefore drives $\ell_X$ in and out of the detectable region during the cycle: at the upper edge (shielding-dominated) is weakened, allowing larger $\epsilon$, while the lower edge (detector-escape dominated) is tightened, excluding smaller $\epsilon$. Thus, modulation primarily reshapes the exclusion where geometric or kinematic thresholds are active, and averages out where the acceptance is already saturated.
Quantitatively, for a $50\%$ modulation in $\epsilon$, we find that the exclusion boundary shifts by up to $\sim 30\%$: the upper edge (corresponding to short-lived mediators) becomes weaker, while the lower edge (corresponding to long-lived mediators) tightens correspondingly. In the region where the static SHiP limit already applies, the modulation effect is small, producing less than a $5\%$ deviation in $\epsilon$.

\begin{figure}[htb!]
    \centering
    \includegraphics[width=0.8\linewidth]{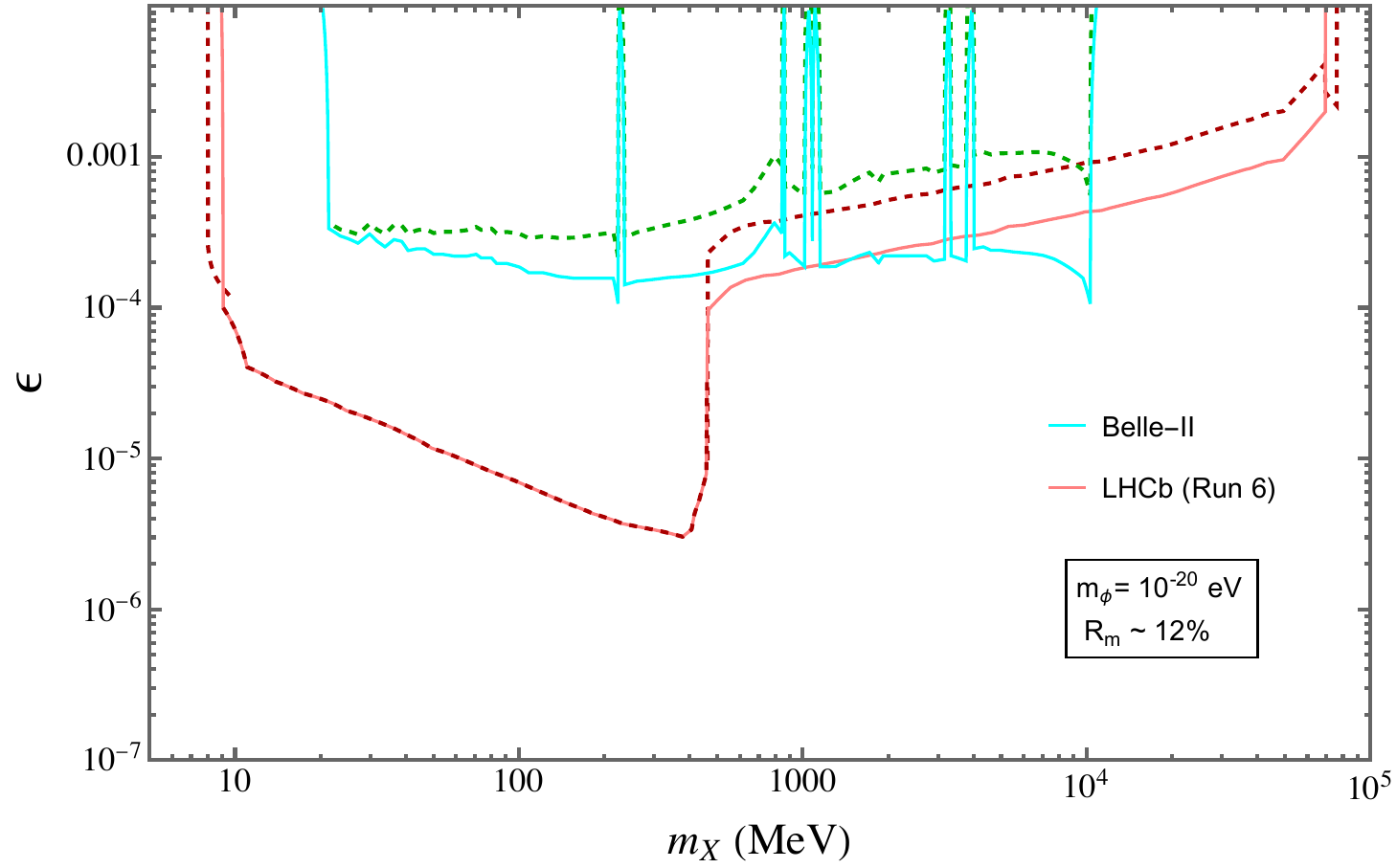}
    \caption{Oscillating resonance effects on future collider sensitivities on the dark photon parameter space. The solid lines denote the standard projections, with the red line signifying the Run-6 projections from inclusive searches of LHCb~\cite{Craik:2022riw}, and cyan for the $50\ {\rm ab^{-1}}$ Belle-II~\cite{Belle-II:2018jsg} projection. The dashed lines correspond to the weakening of the limits in the oscillating DM background.}
    \label{fig:fut}
\end{figure}

Fig.~\ref{fig:fut} illustrates the oscillation effects on future collider sensitivities in the dark photon parameter space, combining both prompt and displaced search strategies. The solid curves denote the standard (static) projected limits, while the dashed curves show the corresponding limits in the presence of an oscillating mediator mass with an optimal modulation ratio $\mathcal{R}_m \simeq 12\%$. The solid red line correspond to LHCb Run-6 inclusive projections~\cite{Craik:2022riw}, which incorporate both prompt and displaced dilepton searches, while the cyan curve shows the projected Belle-II sensitivity~\cite{Belle-II:2018jsg} at $50~{\rm ab^{-1}}$ luminosity.

For $\epsilon \gtrsim 10^{-4}$, the dominant effect arises in the prompt search region, where the exclusion is driven by resonance-based invariant mass analyses. When the modulation amplitude exceeds the detector mass resolution, the resonance is diluted over multiple mass bins, thus weakening the exclusion. This effect is maximised near the optimal modulation ratio and explains the upward shift of the dashed contours relative to the static limits in the prompt-dominated regions.

In contrast, the displaced-vertex region of the LHCb sensitivity ($\epsilon \lesssim 10^{-4}$) is negligibly affected by mass modulation. As discussed earlier, in this regime, the total signal yield is largely insensitive to the mediator mass due to the cancellation between production and decay probabilities and therefore, the solid red and the dashed brown lines coincide in the figure.

The near-threshold behaviour is also visible: for mediator masses close to the kinematic threshold, mass modulation allows the decay channel to open during part of the oscillation cycle, extending sensitivity slightly below the static threshold mass. This effect is subleading compared to the resonance dilution effect but is included consistently in the projected contours.

\section{Peak Reconstruction strategy from events in the oscillating background}\label{sec:reconstruct}

In this section, we present a strategy to reconstruct the central invariant-mass peak shape from a mock invariant mass-binned dataset, a faithful proxy for realistic collider data. We further explore how access to time-binned information can enhance the analysis, clarifying which features are fundamentally inaccessible to invariant mass data alone and which become recoverable once the time-binned information is included.

\subsection{Using mass-binned data}

As discussed in Section~\ref{sec:collider}, an oscillating DM background stretches the invariant mass distribution of the mediator across the full mass modulation window. While the total event yield is conserved, these events are redistributed over a wider mass range, effectively diluting the signal in any given bin and reducing the apparent signal strength.

In a standard mass-scan analysis, the signal is identified by locating the mass bin with the largest excess of events. However, in the presence of mass modulation, this maximum typically occurs near the edges of the modulation window rather than at the central, unmodulated mass value, as illustrated in Fig.~\ref{fig:comphisto}. Consequently, the inferred mediator mass can be biased away from its true value, leading to a potential misidentification of the signal and a corresponding misinterpretation of experimental constraints.

\begin{figure}[tb!]
    \centering
    \includegraphics[width=0.49\textwidth]{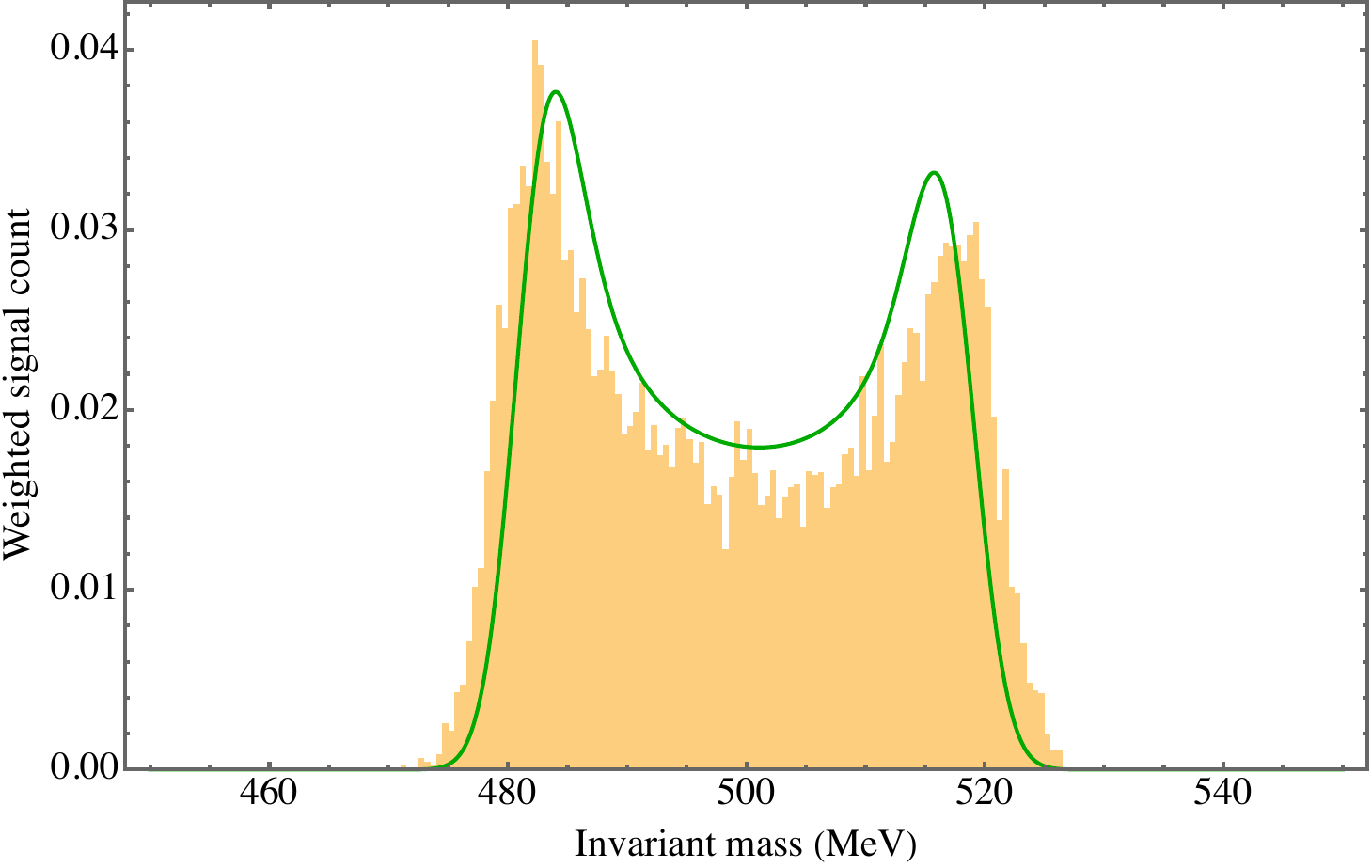}\hfill
    \includegraphics[width=0.49\textwidth]{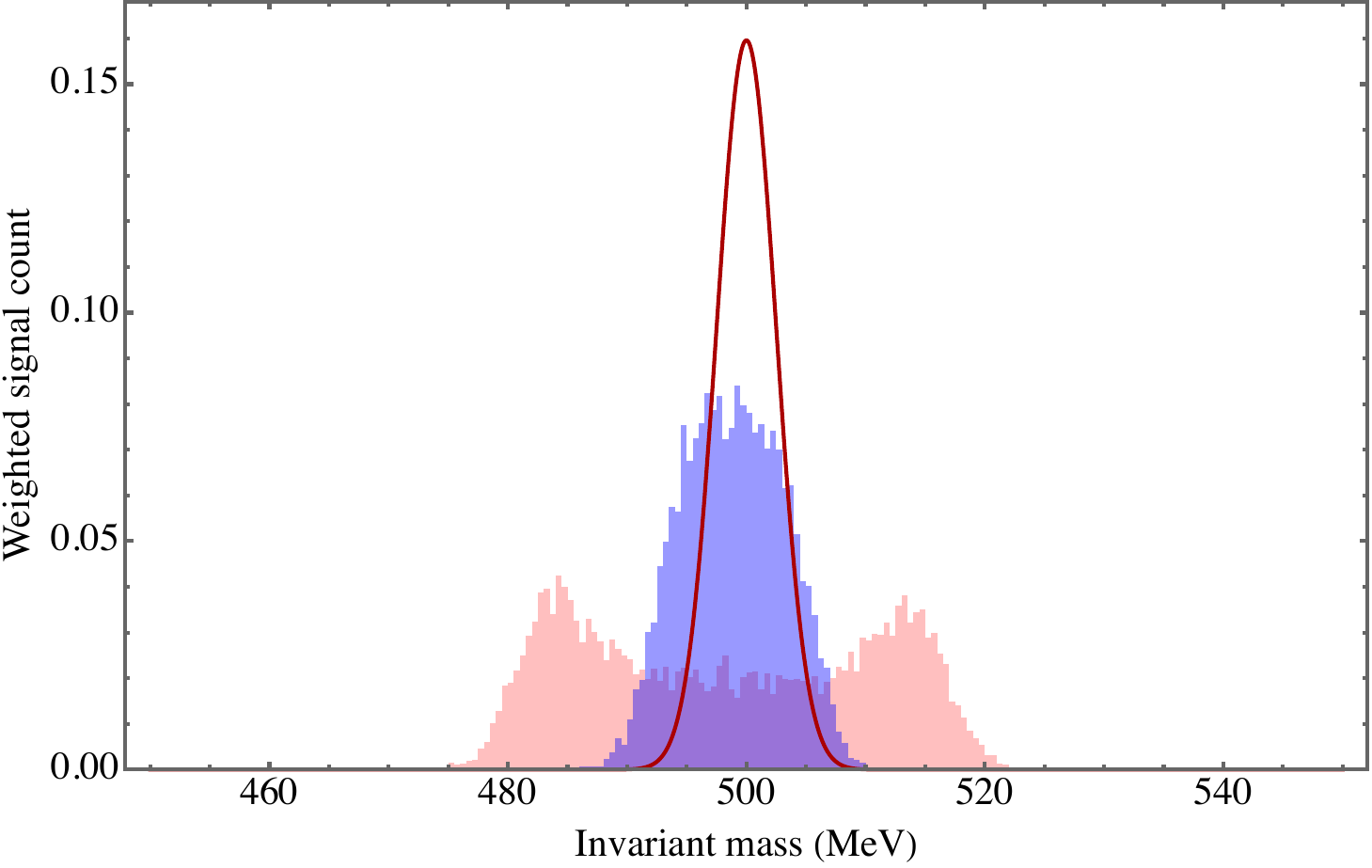}
    \caption{{\bf{\it Left:}} The yellow histogram shows invariant mass distribution for the mock data of $10^4$ events in an ULDM background with $m_0^{\rm true}=500\ {\rm MeV}$ and $\delta m^{\rm true}=20\ {\rm MeV}$. The green curve corresponds to the envelope of the reconstructed distribution obtained by the {\it peak-finding algorithm}, yielding $m_0^{\rm fit}=499.92\pm 0.5\ {\rm MeV}$ and $\delta m^{\rm fit}=17.88\pm 0.625\ {\rm  MeV}$. {\bf{\it Right:}} The pink histogram shows the reconstructed invariant mass distribution obtained by applying the {\it peak-finding} algorithm to the same mock dataset as the left panel. The blue histogram shows the reconstructed distribution obtained after applying the {\it peak-merging} step to the pink reconstruction, yielding a single apparent peak in the regime where the double-peak structure is no longer resolvable. The red curve shows the distribution expected for a static mass hypothesis, centered around $m_0^{\rm true}=500\ {\rm MeV}$ and convoluted only with a finite detector resolution ($\Delta m_{\rm res}=2.5\ {\rm MeV}$), which is plotted for only shape comparison. All distributions are normalised as probability density functions.}
    \label{fig:massbin}
\end{figure}

In the {\bf\it Left} panel of Fig.~\ref{fig:massbin}, we compare invariant mass distributions for a mock dataset of $10^4$ events in the presence of an oscillating DM background. The true central dark photon mass is $m_0^{\rm true}=500~{\rm MeV}$, with a mass modulation amplitude of $\delta m^{\rm true}=20~{\rm MeV}$. We design the reconstruction algorithm to operate in two stages: {\it peak-finding} and {\it peak-merging}. In the {\it peak-finding} step, the two most prominent peaks near the edges of the modulation window are identified, allowing an initial estimate of the modulation amplitude $\delta m$. The subsequent {\it peak-merging} step iteratively narrows the mass window, effectively collapsing the modulated structure until the two peaks become unresolvable. This procedure enables a reconstruction of the underlying, unmodulated invariant mass peak shape.

When the modulation window becomes large, events get significantly diluted across bins, resulting in a correspondingly reduced signal-to-noise ratio (SNR). We analyse the histogram by locating local maxima in the weighted event counts and selecting the two most significant peaks that satisfy both a minimum separation criterion and a threshold bin height. These requirements ensure that the identified peaks are not statistical fluctuations and that the modulation structure is sufficiently well-resolved. 

This selection is essential to the validity of the method, as the modulation can be reliably modeled only when the two dominant peaks at the edges of the modulation window are clearly resolved. The detected peaks are interpreted as the boundaries of the modulation envelope. The central mass is then estimated as $m_0^{\rm est}$, defined by the midpoint between the two peaks, while the modulation amplitude $\delta m^{\rm est}$ is taken as half of their separation. 

Setting a threshold at half the height of the tallest bin and requiring a minimum separation of 3 MeV between peaks, our analysis yields estimates of $m_0^{\rm est}=499.92\pm0.5\ {\rm MeV}$ and $\delta m^{\rm est} =17.88\pm 0.625\ {\rm MeV}$. The uncertainty estimates are obtained from an ensemble of 1000 pseudo-experiments, in which the {\it peak-finding} step is applied independently to each realisation. The resulting estimates correspond to the green envelope as shown in the {\it left} panel of Fig.~\ref{fig:massbin}.

Without time-binned data, the modulation phase cannot be accurately determined. However, we can estimate the phase using $\theta=\cos^{-1}\left((m_X-m_0^{\rm est})/\delta m^{\rm est}\right)$ for the two identified peaks and confirm that they occur around $\theta=0$ and $\pi$, as expected for modulation extrema.

The {\it peak-merging} algorithm tracks how the number of resolvable peaks evolves as the modulation window is progressively reduced. Once an estimate of the central mass, $m_0^{\rm fit}$, is obtained from the {\it peak-finding} step, we generate a sequence of mock invariant mass histograms with decreasing modulation amplitudes $\delta m$. For each $\delta m$, we count the number of statistically significant local maxima in the weighted mass distribution. 

The algorithm scans for the point where the two peaks associated with the modulation edge merge into a single, broadened structure. This merging point effectively marks the effective resolution threshold, beyond which the detector can no longer resolve the two modulated peaks. Physically, it corresponds to the regime where the modulation-induced broadening becomes comparable to the experimental mass resolution.

In our simulation, the merging point is defined as the smallest modulation amplitude $\Delta m_{\rm est}$, for which the {\it peak-merging} algorithm identifies a single statistically significant peak across five consecutive values of $\delta m $, ensuring robustness against bin-to-bin fluctuations. From an ensemble of 1000 pseudo-experiments, the limiting modulation amplitude is estimated to be $\Delta m_{\rm est} = 4.83\ {\rm MeV}\pm 0.5\ {\rm MeV}$, which is slightly larger than the nominal mass resolution for a 500 MeV mediator in current searches, $\Delta m_{\rm res}\simeq 2.5\ {\rm MeV}$ for Belle-II~\cite{Belle-II:2024wtd} and $\sim$ 3.5 MeV as interpolated from LHCb detector performance~\cite{lhcb}. 

We sample the oscillation phase from a uniform distribution over ($0,2\pi$), assigning an independent phase to each event. Assuming a cosine modulation of the mass, we construct the smeared invariant mass distributions. In the {\it right} panel of Fig.~\ref{fig:massbin}, the pink histogram shows the double-peak structure obtained from the {\it peak-finding} algorithm. The blue curve represents the reconstructed single-peak distribution from {\it peak-merging} procedure, using $m_0^{\rm est} = 499.92\ {\rm MeV}$ and a Gaussian convolution width $\Delta m_{\rm est} = 4.25\ {\rm MeV}$. For shape comparison, the red curve shows the static (unmodulated) case, corresponding to a central peak at $m_0^{\rm true}=500\ {\rm MeV}$, smeared with the detector resolution $\Delta m_{\rm res}=2.5\ {\rm MeV}$.

While the shape of the central resonance peak can be accurately recovered by de-modulating and re-weighting the data using the inferred modulation parameters, this procedure can't reproduce the correct absolute normalisation (unless the total number of signal events is known). In the simulation, the total number of generated signal events is known by construction and can be imposed externally, enabling a controlled comparison between reconstructed and reference histograms. Importantly, this total yield is not recovered by the {\it peak-finding} or {\it peak-merging} algorithms, but supplied as an external input. Therefore, in a real experimental setting, however, the total event yield cannot be reliably inferred from the modulated invariant-mass distribution alone. The modulation weights suppress the observed signal, and additional effects such as finite detector resolution and the limited mass integration range further complicate the overall normalisation.

As a result, while features such as the peak shape, modulation phase, and relative distortions can still be extracted for qualitative spectral analysis, any inference that relies on the absolute signal strength, such as cross-section determination, coupling constraints, or comparisons with background models, requires access to time-resolved event information rather than mass-binned data alone. This limitation highlights the importance of time-binned information in analyses aiming to quantify the absolute impact of oscillating DM backgrounds on traditional bump-hunt searches. In the next subsection, we present a prescription for recovering the exact peak normalisation using time-domain data.

\subsection{Using time-binned data}

In this section, we illustrate how phase information can be extracted from time-binned data and used to reconstruct both the shape and the absolute normalisation of the central invariant mass peak. Unlike analyses based solely on mass-binned information, access to event time stamps enables a direct fit to the underlying modulation pattern, allowing the signal yield to be recovered alongside the modulation parameters.

A typical caveat of this approach is that the reconstruction relies on assuming a specific functional form for the modulation, most commonly a sinusoidal variation of the mediator mass induced by an oscillating DM background. This assumption constitutes a modeling choice rather than a first-principles necessity. If the true dynamics of the DM field deviate from simple harmonic behaviour--for example, due to anharmonic motion, multi-frequency components, or transient effects--residual distortions may remain in the reconstructed signal. Nevertheless, within the regime where the modulation is approximately periodic, time-binned analyses provide a powerful and complementary handle on both the spectral shape and absolute normalisation of the signal.

\begin{figure}[tb!]
    \centering
    \includegraphics[width=0.5\textwidth]{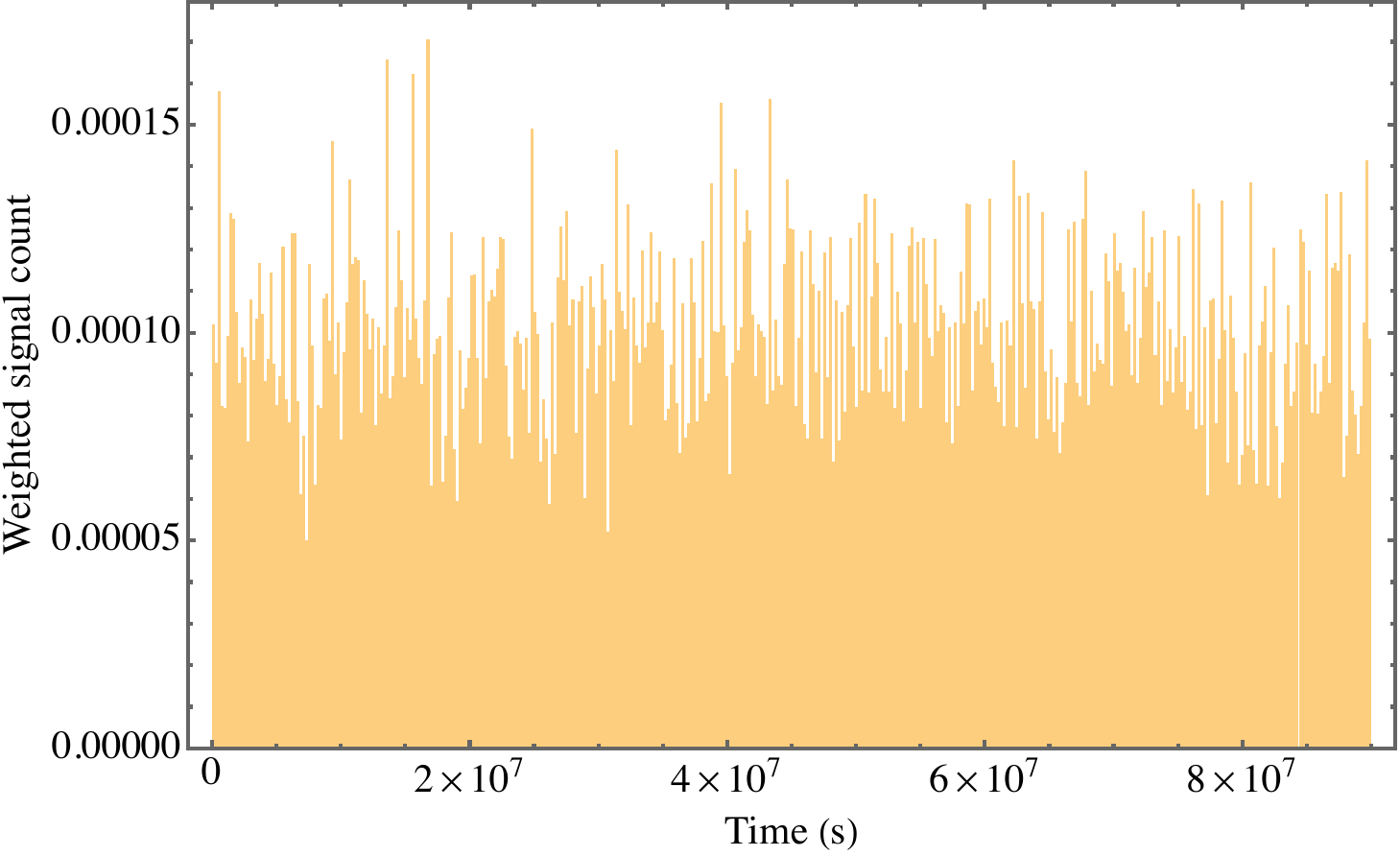}\hfill
    \includegraphics[width=0.49\textwidth]{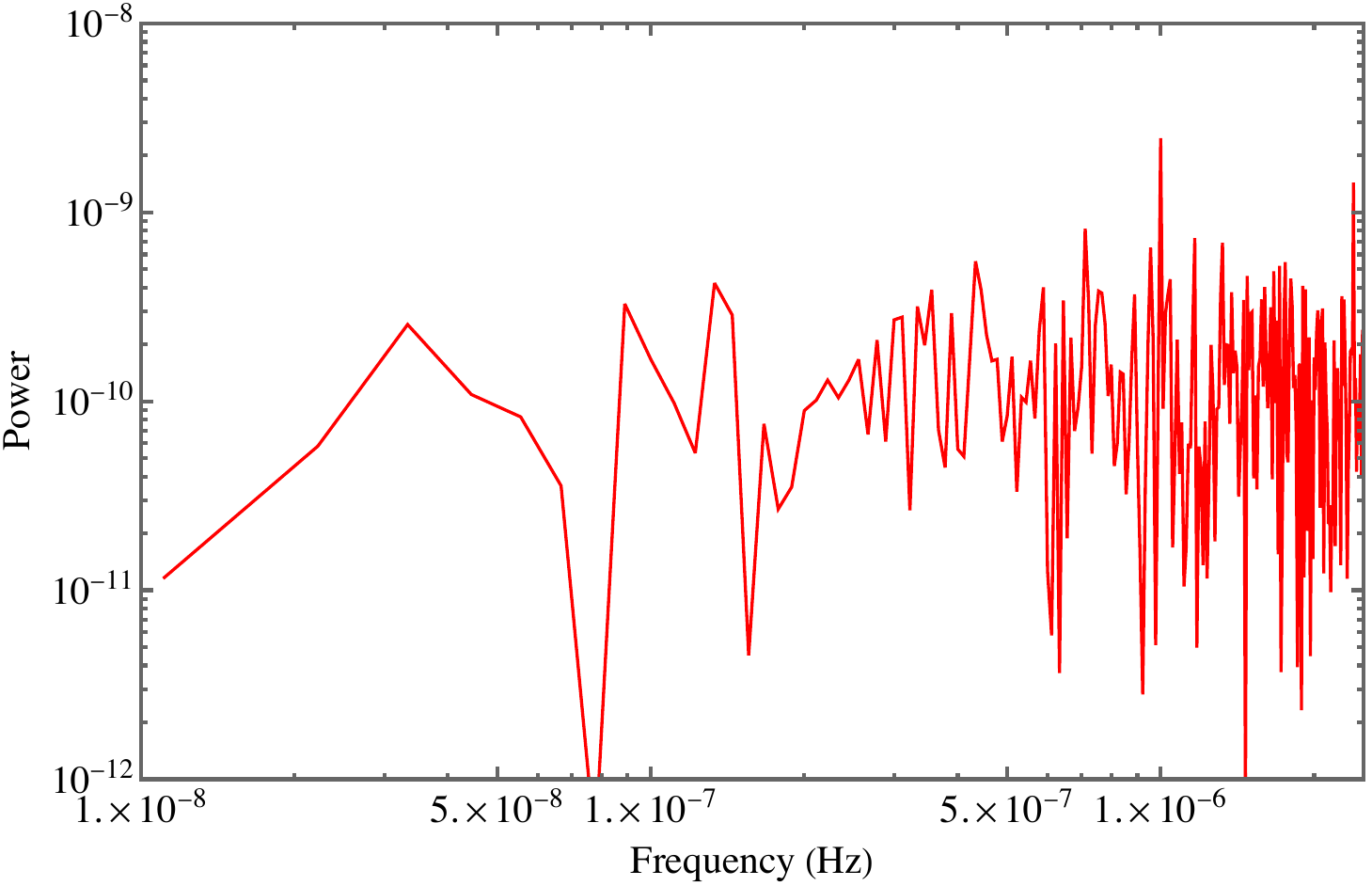}
    \caption{{\bf {\it Left\:}} Time-binned weighted event counts for a mock dataset with $\delta m^{\rm true} = 20~{\rm MeV}$ and $10^4$ signal events distributed uniformly over a 3-year run. The weights scale as $1/m_X^2(t)$, producing a time-dependent modulation that is not visually apparent due to statistical fluctuations. {\bf {\it Right:}} FFT power spectrum of the same dataset, showing a clear peak at the injected modulation frequency corresponding to $m_\phi \sim 10^{-20}~{\rm eV}$ ($\omega=10^{-6}\ {\rm Hz}$), demonstrating that the periodic signal can be recovered in the frequency domain even when it is obscured in the time series.} 
    \label{fig:time-binned}
\end{figure}

In Fig.~\ref{fig:time-binned}~({\it left}), we show a mock time-binned distribution of weighted event counts for a dark photon with true central mass $m_0^{\rm true} = 500~{\rm MeV}$ and modulation amplitude $\delta m^{\rm true} = 20~{\rm MeV}$. The events are sampled uniformly over a 3-year run with $10^4$ signal events, and the weights reflect the time-dependent production cross-section, which scales as $1/m_X^2(t)$. Although the modulation is not immediately apparent in the time domain due to statistical fluctuations, applying a fast Fourier transform (FFT) to the binned data reveals a clear peak in the power spectrum~\cite{Schelfhout_2021}, shown in Fig.~\ref{fig:time-binned}~({\it right}), at a modulation frequency $\omega \sim 10^{-6}~{\rm Hz}$. This frequency corresponds to an ULDM of mass $m_\phi \sim 10^{-20}~{\rm eV}$. 

Once the modulation frequency $\omega$ is identified from the FFT of the time-binned signal, we apply phase folding to coherently amplify the periodic modulation. Each event time $t_i$ is mapped to a phase angle $\theta_i = \omega t_i$, thereby projecting the multi-year dataset onto a single oscillation cycle. This procedure stacks contributions from many oscillation cycles and reveals a sinusoidal modulation in the event rate that would otherwise be obscured by statistical noise. 

The phase-folded data are then binned in $\theta$ and fitted to a model of the form
\begin{align}
    f_{\rm sig}(\theta) = \frac{a}{(1 + \delta m/m_0 \cos(\theta - \phi))^2}
    \label{eq:fsig}
\end{align} 
which follows directly from the assumed time dependence of the effective mediator mass, $m_X(\theta) = m_0 + \delta m \cos(\theta - \phi)$, and the corresponding production rate scales as $\sigma\, \propto\, 1/m_X^2(\theta)$. Under the assumption of negligible background, this functional form captures the full modulation-induced variation of the signal rate.

The fit yields direct estimates of the modulation amplitude ratio $\delta m/m_0$, the phase offset $\phi$, and the overall normalisation $a$. In contrast to the mass-binned analysis, the normalisation is no longer degenerate with the modulation parameters in the zero-background case, allowing the absolute event yield to be reconstructed. The enhancement in SNR arises from coherently folding many oscillation cycles into a single phase profile, in close analogy to matched-filtering techniques~\cite{Panelli_2020} employed in precision timing and frequency-domain searches. It is, however, important to note that the absolute $\delta m$ from the best-fit parameters can only be obtained if $m_0$ is known. In Fig.~\ref{fig:time-binned}~({\it left}), we show an illustrative phase-folded distribution together with the corresponding best-fit phase model for a single pseudo-experiment. From an ensemble of pseudo-experiments, we obtain a best-fit modulation amplitude $\delta m^{\rm fit}= 19.24\pm 2.65\ {\rm MeV}$, using $m_0^{\rm est}=499.92\ {\rm MeV}$ from the mass-binned data.

\begin{figure}[tb!]
    \centering
    \includegraphics[width=0.5\textwidth]{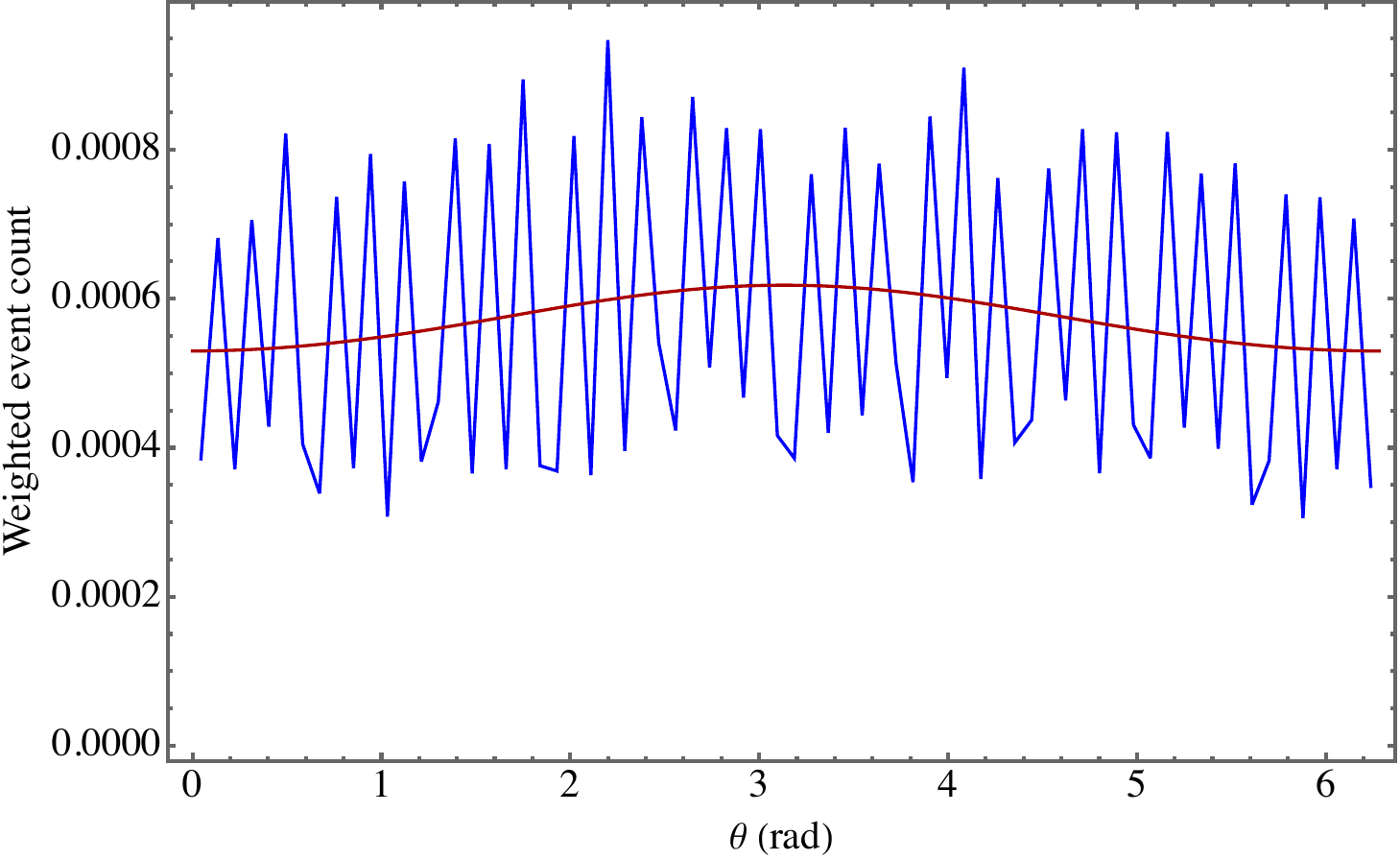}\hfill
    \includegraphics[width=0.49\textwidth]{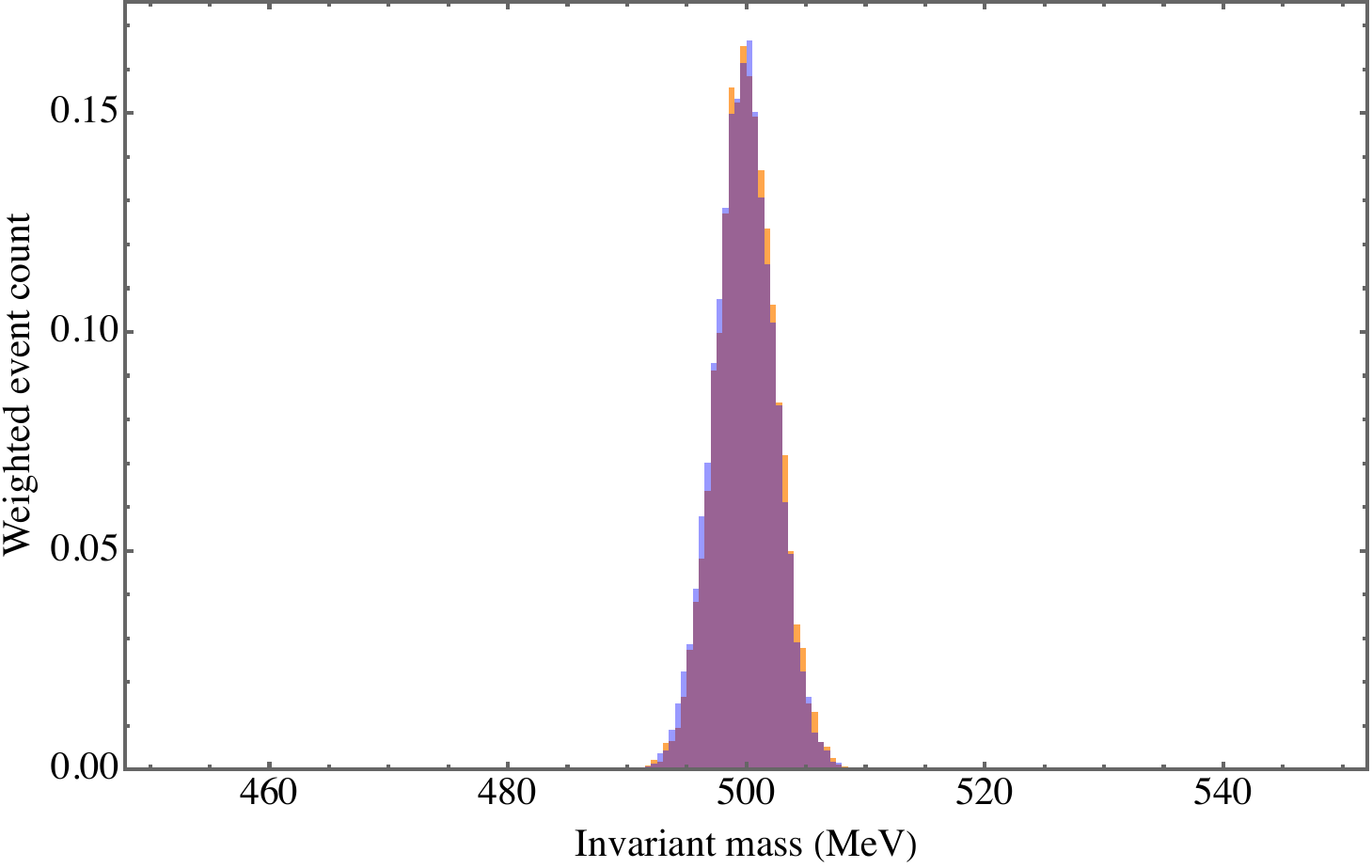}
    \caption{{\bf {\it Left:}} Phase-folded weighted event counts for the zero-background case after mapping event times $t_i$ to phases $\theta_i=\omega t_i$, revealing a coherent sinusoidal modulation over one oscillation cycle. The red curve shows the best-fit modulation model $f_{\rm sig}(\theta)$ with extracted parameters $\omega \sim 10^{-6}\ {\rm Hz}$, $\delta m^{\rm fit}\simeq 19.26\ {\rm MeV}$. {\bf {\it Right:}} Reconstructed invariant mass distribution obtained after de-modulating and re-weighting the phase-folded signal, compared to the reference static case.}
    \label{fig:time-binning reconstruction}
\end{figure}

Following the extraction of modulation parameters from the phase-folded distribution, we reconstruct the unmodulated signal by inverting the time-dependent weights. Using the best-fit modulation amplitude $\delta m^{\rm fit}$, we define an effective time-dependent mass $m_X(t) = m_0^{\rm fit} + \delta m^{\rm fit} \cos(\omega t)$ and compute weights proportional to $1/m_X^2(t)$, which model the modulation-induced suppression of the production rate. These weights are then inverted and applied to the original weighted, time-binned event counts to recover the underlying, unmodulated event yield.

As shown in Fig.~\ref{fig:time-binning reconstruction}~({\it right}), this procedure successfully reconstructs the central resonance peak. The blue histogram denotes the reconstructed unmodulated distribution obtained after inverting the time-dependent weights, while the orange histogram shows a static reference sample drawn from a Gaussian distribution centred at $m_0^{\rm true}=500\ {\rm MeV}$ with resolution $\Delta m_{\rm res}=2.5\ {\rm MeV}$. The excellent agreement in total event yield confirms that the unweighting procedure accurately recovers the unmodulated signal under the assumed modulation model.

\begin{figure}[htb!]
    \centering
    \includegraphics[width=0.49\textwidth]{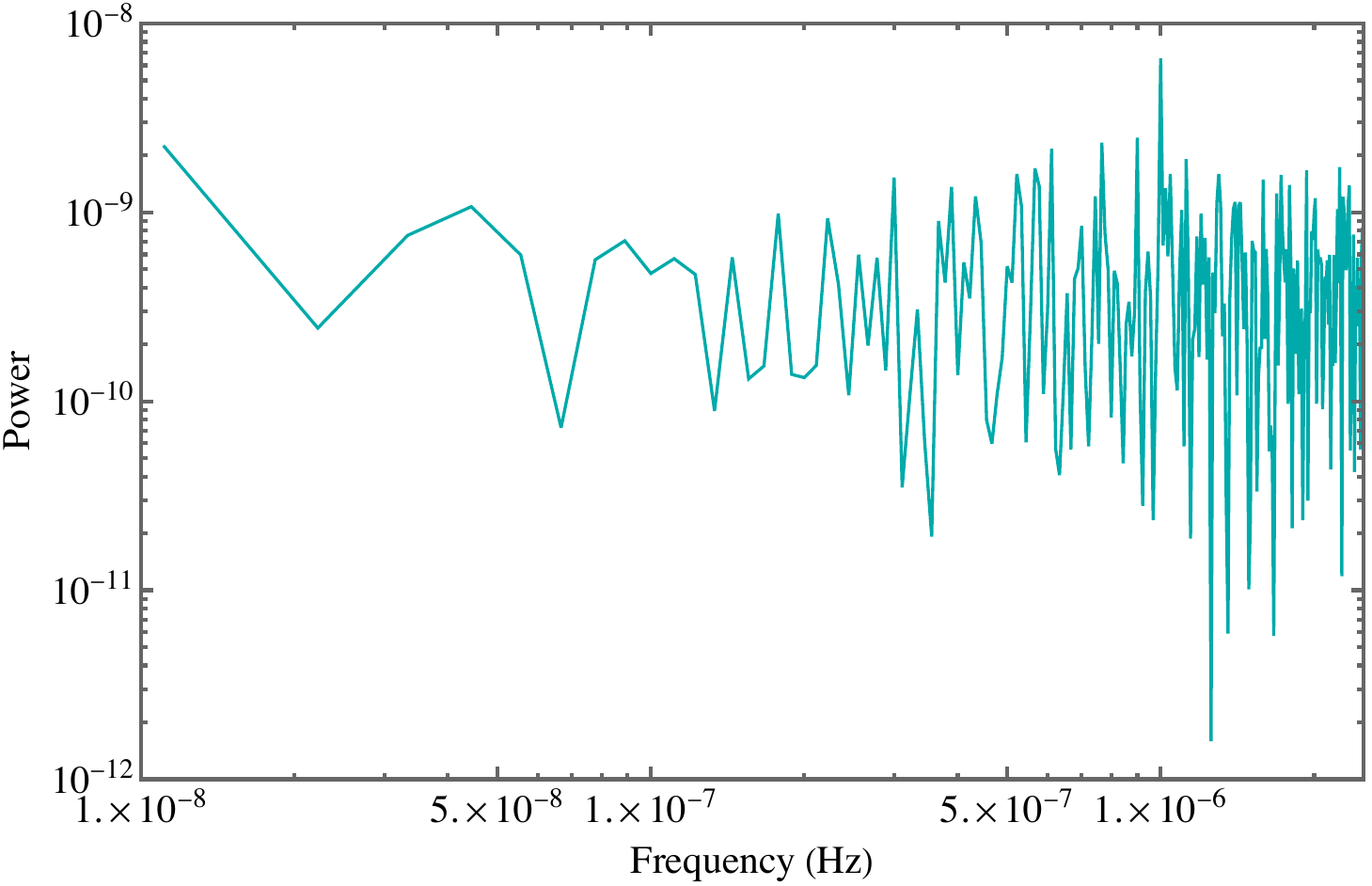}\hfill
    \includegraphics[width=0.5\textwidth]{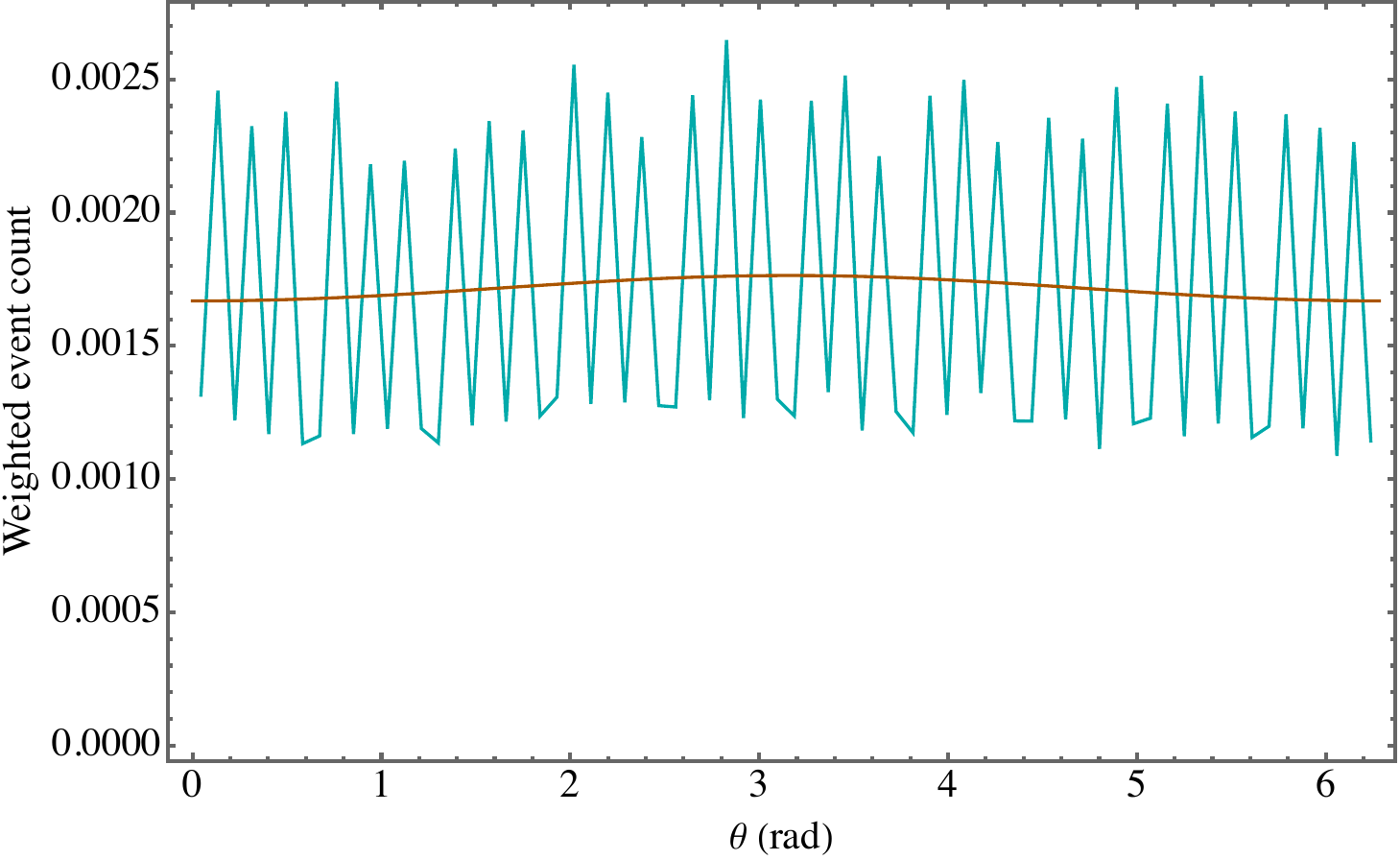}
    \caption{{\bf {\it Left:}} FFT power spectrum of the time-binned weighted event counts in the presence of a DC background with $S/B \approx 1/2$. The modulation peak at the true signal frequency remains visible but sits on an elevated noise floor compared to the zero-background case, reducing its statistical significance. {\bf {\it Right:}} Phase-folded weighted event counts after mapping event times to phases using the extracted modulation frequency. The coherent oscillatory structure is distorted by the additive background, leading to a reduced modulation depth in the fit-function. The orange curve shows the best-fit phase model $f_{\rm obs}(\theta)$ with extracted parameters $\omega \sim 10^{-6}\ {\rm Hz}$, $\delta m^{\rm fit}\simeq 8.1\ {\rm MeV}$.}
    \label{fig:time-binning reconstruction-bkg}
\end{figure}

Finally, we comment on the impact of a finite, time-independent (DC) background on the time-domain analysis. The presence of background events introduces additional shot noise in the time-binned signal, which manifests in the frequency domain as an elevated broadband noise floor in the FFT power spectrum. This is depicted in Fig~\ref{fig:time-binning reconstruction-bkg}~({\it left}). As the background level increases, the contrast between the modulation peak and the surrounding noise is progressively reduced. Beyond a certain point, the modulation peak can no longer be reliably distinguished from statistical fluctuations, rendering both the frequency identification and subsequent phase folding ineffective.

\begin{figure}[tb!]
    \centering
    \includegraphics[width=0.49\textwidth]{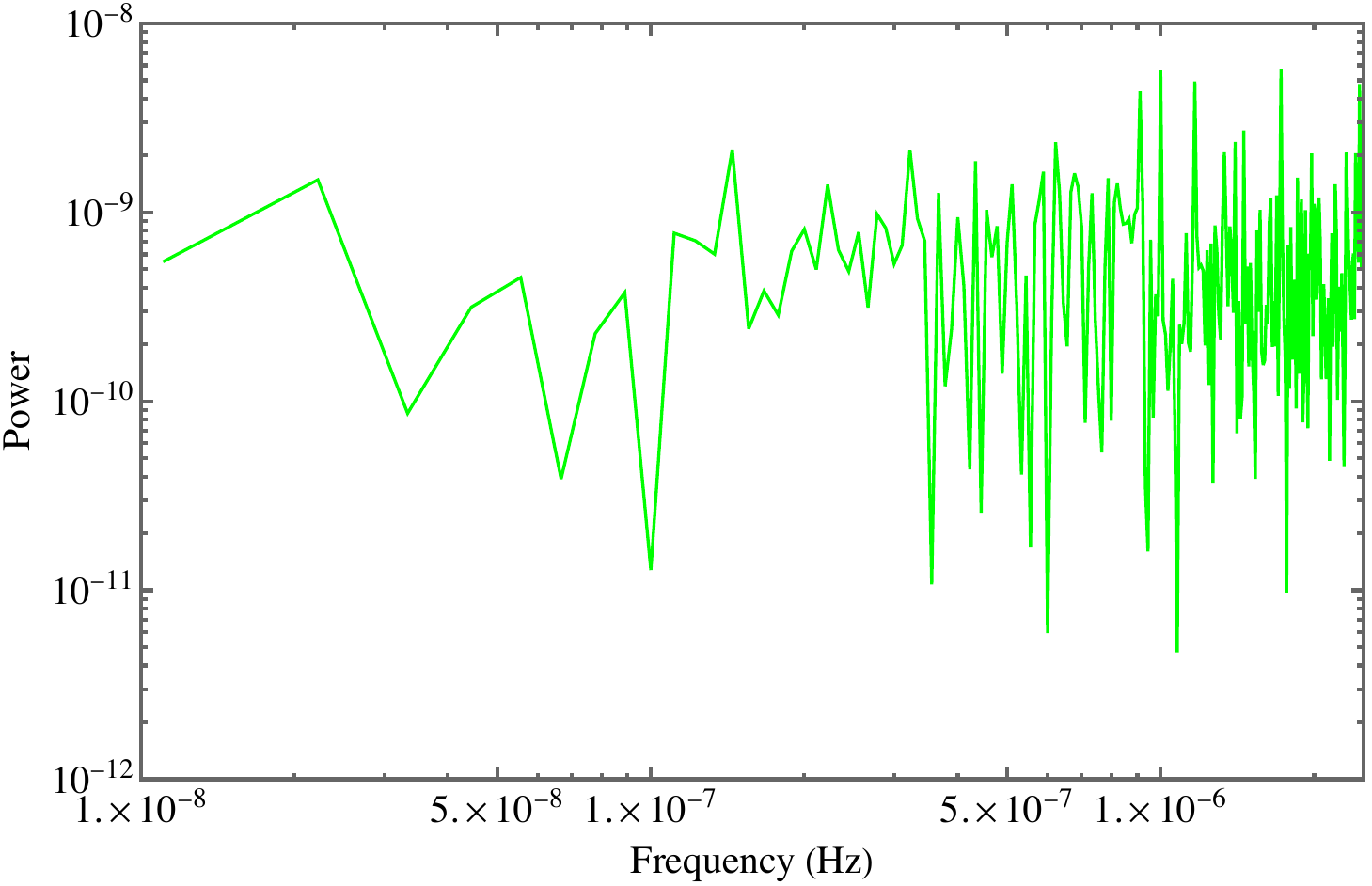}\hfill
    \includegraphics[width=0.5\textwidth]{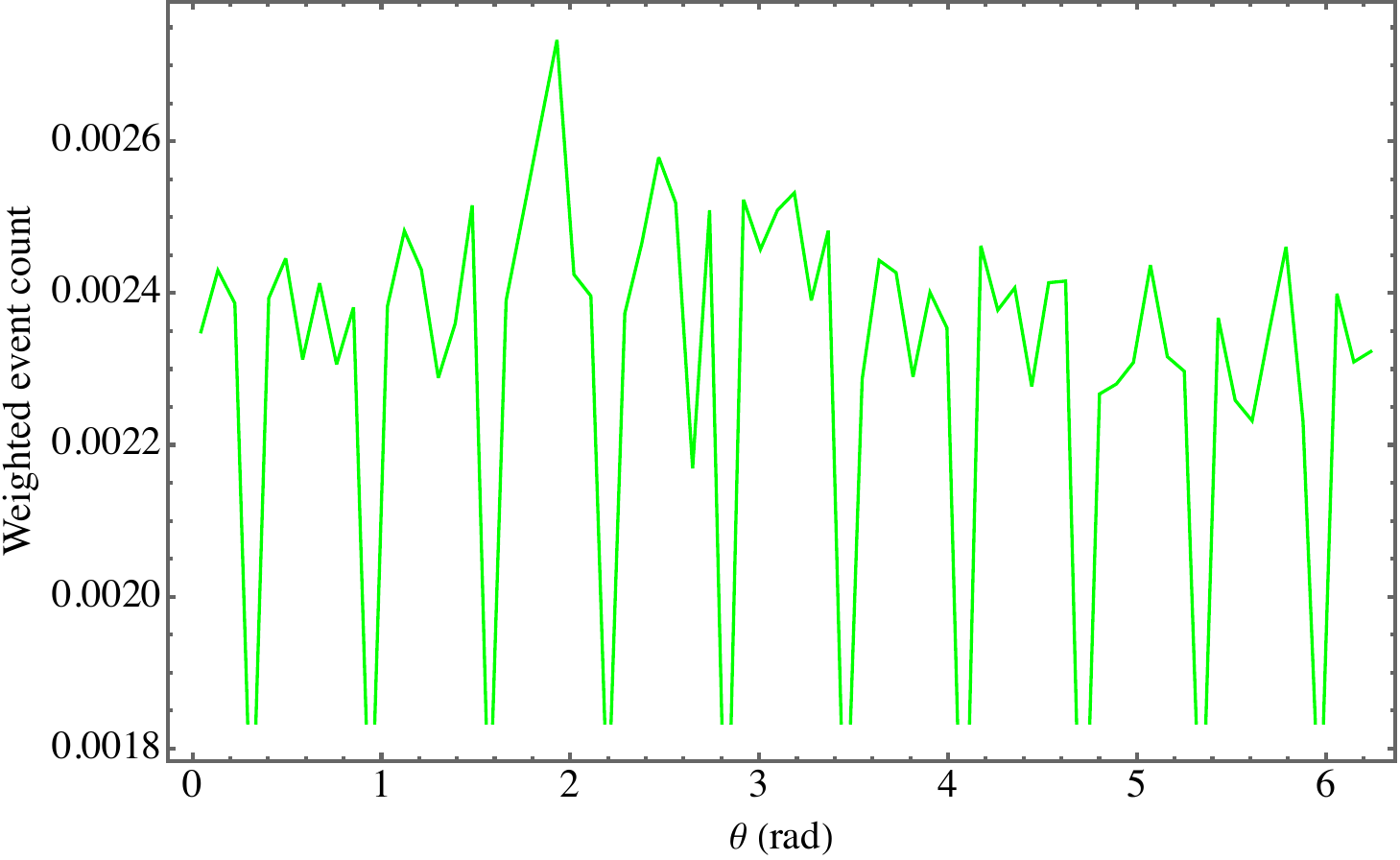}
    \caption{{\bf {\it Left:}} FFT power spectrum of the time-binned weighted event counts in the presence of a larger DC background ( $S/B \approx 1/3$) where the noise floor hides the signal peak. FFT algorithm mis-identifies the peak frequency as $\omega\sim 3.88\times 10^{-6}\ {\rm Hz}$. {\bf {\it Right:}} Phase-folded weighted event counts after mapping event times to phases using the extracted modulation frequency. The oscillatory structure is no-longer visible, rendering the fitting analysis redundant.}
    \label{fig:time-binning reconstruction-Largebkg}
\end{figure}
To quantify the impact of a time-independent background on the time-domain analysis, we fix the total number of events in the mock dataset to $10^4$ and progressively increase the fraction of background events. In this setup, we find that the modulation peak in the FFT power spectrum remains clearly identifiable down to a signal-to-background ratio of approximately $ S/B\approx1/3$. Below this level, the modulation peak becomes comparable to stochastic fluctuations in the broadband noise floor, and its statistical significance rapidly degrades. This transition therefore, marks the practical limit at which the oscillation frequency can be robustly extracted using frequency-domain methods alone in our mock analysis. This is explicitly illustrated in Fig.~\ref{fig:time-binning reconstruction-Largebkg}. However, we emphasise that the precise value of this threshold is not universal and depends on the total statistics, observation time, and binning choices; it should be regarded as illustrative for the present setup.

Even in regimes where the FFT peak remains marginally detectable, the presence of a sizable DC background significantly distorts the phase-folded distribution. Since the background contributes an additive, phase-independent component, fitting the folded data with a signal-only phase model leads to a systematic suppression of the inferred modulation amplitude. The observed phase-folded distribution in the presence of background is instead given by
\begin{align}
    f_{\rm obs}(\theta)=a\left((1-\lambda)f_{\rm sig}(\theta)+\lambda\right)
\end{align}
where $\lambda$ denotes the fractional contribution of the DC background and $f_{\rm sig}$ is given by ~\eqref{eq:fsig}. As $\lambda$ increases, the modulation depth for the phase-folded signal is reduced, biasing the fitted modulation amplitude toward smaller values and degrading the precision of parameter extraction.

In this background-dominated regime, the distribution of fitted modulation amplitudes obtained from pseudo-experiments becomes increasingly non-Gaussian, reflecting statistical fluctuations and the fact that the likelihood becomes weakly sensitive to the modulation amplitude. To characterise the overall degradation of the fit, we therefore combine the statistical spread of the estimator with its bias relative to the injected value into an effective uncertainty, $\sigma^{\rm eff}$, where $\sigma^{\rm eff}=\sqrt{\sigma_{68}^2+b^2}$. $\sigma_{68}$ denotes half the width of the central 68\% interval of the fitted $\delta m$ distribution obtained from the pseudo-experiments and $b$ is the bias of the median $\delta m^{\rm fit}$ relative to the injected value $\delta m^{\rm true}$. This provides a more faithful measure of sensitivity loss than the interval width alone.

Tab.~\ref{tab:background_bias} summarises the best-fit modulation amplitudes with $\sigma^{\rm eff}$ estimates obtained for different signal-to-background ratios. The rapid growth of $\sigma^{\rm eff}$ with decreasing $S/B$ ratio indicates that, while the phase-based analysis continues to return a best-fit value of $\delta m$, the measurement loses statistical constraining power once $S/B\lesssim 1$. In the presence of a substantial DC background, the phase-dependent modulation is diluted by a factor $(1-\lambda)$, leading to a systematic suppression of the inferred modulation amplitude $\delta m$. As a result, the phase-folded data no longer determine a unique modulation amplitude, but instead only constrain the maximum $\delta m$ compatible with the observed modulation depth. In this regime, the fit result should therefore be interpreted as an upper bound rather than a precise measurement.
\begin{table}[t]
\centering
\begin{tabular}{|c  |c|}
\hline
$S/B$ & $\delta m^{\rm fit}\pm \sigma^{\rm eff}$ (MeV)  \\
\hline
3   & $15.98\,\pm\, 5.65$  \\
2   & $14.64\,\pm\, 6.56$  \\
1   & $10.52\,\pm\, 11.39$   \\
0.5 & $7.84\,\pm\, 13.11$ \\
0.3 & $4.99\,\pm\, 15.39$ \\
\hline
\end{tabular}
\caption{Best-fit modulation amplitudes and corresponding uncertainty estimates as a function of signal-to-background ratio.}
\label{tab:background_bias}
\end{table}
Finally, the unweighting procedure used to reconstruct the unmodulated invariant mass peak ceases to be valid in the presence of an unknown DC background. Since the signal modulation enters multiplicatively through the time-dependent weights, while the background contributes additively, inverting the signal weights on the combined (signal+background) distribution does not recover the true unmodulated signal. Consequently, without independent control over the background contribution, time-domain unweighting can no longer reproduce the correct peak normalisation.

\section{Conclusions}
\label{sec:conclusions}

The interactions of ultralight dark matter with Standard Model fields are typically described using effective operators. We discuss UV completions of these operators in which dark matter interactions with electrons and photons are loop-suppressed. In this case, searches for the mediator at colliders can be an alternative search strategy to measuring the variation of fundamental constants. The oscillations of the dark matter field induce oscillating mass terms or couplings of the mediator, which lead to ``oscillating resonances" that smear out the resonance peak. The frequency of this oscillation is related to the dark matter mass and it is remarkable that the oscillation period is within seconds to years for dark matter masses $10^{-22} \,\text{eV}\lesssim m_\phi \lesssim10^{-15}\,\text{eV} $, which are timescales short enough to be probed at colliders and long enough not to be within the detector resolution.  

We discuss different models, distinguishing scalar and vector mediators and consider Yukawa and gauge couplings to SM particles as well as kinetic mixing. While these models all result in different effective Wilson coefficients and therefore predictions for the variation of the fine-structure constant and the electron mass, a resonance with oscillating mass is a novel collider signature common to all the models we consider.

At the example of a spin-1 mediator with kinetic mixing, we calculate the sensitivity of searches for resonances decaying into muons at Belle-II and LHCb to an oscillating resonance. We model the expected signal by modulating a resonance peak, and average over a full oscillation cycle. Instead of a narrow resonance for which most all signal events are expected in a single mass bin, the signal from an oscillating resonances is smeared out over several mass bins. A standard bump-hunt can only detect a significantly reduced `resonance' towards the low mass end of the oscillation window. Depending on the mass and interactions of the ultralight dark matter, we show that existing constraints on the kinetic mixing parameter from LHCb are weaker by a factor 4-5. Another interesting effect of an oscillating resonance mass is that decay thresholds can be passed periodically. A resonance with a mass close but below $2m_\mu$ can decay into two muons whenever the dark matter oscillations induce a slightly larger mass. As a result, resonance searches for this final state can constrain such a state even for $m_X\lesssim 2m_\mu$. 

In contrast to searches for prompt decays at colliders, searches for displaced vertices at beam dumps are largely insensitive to mass oscillations, because the mass dependence of the decay length and the cross section cancel. Models in which the coupling parameter of the mediator oscillates instead do lead to oscillating decay lengths, which allow for the mediator to decay inside the acceptance range even if it would not in the absence of the dark matter. This increases the sensitivity of  beamdump searches to oscillating resonances compared to the case of a time-independent signal. We show the projected range of dark photon searches at ShiP, for which the effect is within a factor of 2 of the projected sensitivity to a generic long-lived particle.

Standard bump-hunt searches are not the optimal strategy to discover this novel signal. The periodic oscillation induces a modulation of the resonance line shape, resulting in an asymmetric double-peaked profile in the invariant mass spectrum. We present an algorithm to unfold this modulation if both peaks of the profile can be detected in mass bins at the edges of the oscillation window. For different model parameters, we derive a minimum separation criterion and a threshold event count that is compatible with ultralight dark matter models and show how the peak-finding and peak-merging algorithms can recover this signal, but are ultimately limited in determining the total cross section due to the loss of information by averaging over the oscillation period. A misidentification of the second peak in the invariant mass spectrum would result in an interpretation as a wide resonance, smaller and at a lower invariant mass than the actual mass parameter of the oscillating resonance.    

If time-stamped data is available it is possible to reconstruct the oscillating modulation of the resonance using a fast Fourier transform on the time-binned data. We show how the signal peak can be reconstructed in the frequency domain for different signal-to-noise ratios. As long as the signal is strong enough, our strategy can reliably find the resonance. If the signal is below the statistical background it is challenging to extract the resonance peak and for signal-to-noise ratios $S/B\leq 1$, the sensitivity to an oscillating signal is below one standard deviation.

In contrast to resonance searches for heavy mediator states, the discovery of a resonance  alone wouldn't be sufficient to determine the nature of dark matter, discovering an `oscillating resonance' at a collider would be a smoking gun signal for ultralight dark matter. It is compelling that even for all the models discussed here, the sensitivity of colliders is on the level of precision experiments searching for variations in fundamental constants, making a discovery at a collider a realistic possibility.

\section*{Acknowledgements}

We acknowledge support from the UKRI Future Leader Fellowship DarkMAP (Ref. no. MR/Y034112/1). SC thanks Jonathan Feng and Maurizio Giannotti for initial discussions. This article is based upon work from COST Action COSMIC WISPers CA21106, supported by COST (European Cooperation in Science and Technology)
\appendix

\section{Loop Integrals}
The parameter integrals for the loop-induced contributions~\eqref{eq:scalar}, \eqref{eq:Xmass} and \eqref{eq:XFS} read 
\begin{align}
    I_S&=\int du\frac{u(1+u)}{u+\tau(1-u)^2}\,, \\
    I_X&=\int du\frac{u(2-u)}{u+\tau(1-u)^2}\,,\\
    I_\text{FS}&=3\int du\, u\left[(3u-4)\log(u+\tau(1-u)^2)+ \frac{\tau(1-u)^3}{u+\tau(1-u)^2}\right]\,.
\end{align}
Our expression for $I_X$ agrees with the result in \cite{Batell:2009yf} in the limit $m_\phi =0$. 

\section{Signal fraction}
\label{sec:frac}

For the decay to be kinematically feasible,
\begin{align}
    m_X(t)& =m_X^0+\delta m \cos\omega t > m_{\rm th}\,,
\end{align}
which leads to the following condition:
\begin{align}
    \delta m\cos\omega t & > m_{\rm th}-m_X^0\,,\nonumber\\
    \cos\omega t &> \frac{m_{\rm th}-m_X^0}{\delta m}\,,\nonumber\\
\end{align}

upon assuming $\omega t=\theta$ and $x=\frac{m_{\rm th}-m_X^0}{\delta m}$, we need to find the fraction of the time in an oscillation cycle when 
\begin{align*}
    \cos\theta >& x\,, \\
    \theta \in& \left(0,\cos^{-1} x\right) \\
    &\left(2\pi-\cos^{-1}x,2\pi\right)\,.
\end{align*}
Therefore, the fraction is 
\begin{align}
    f_{\rm ON}&\equiv \frac{2\cos^{-1}x}{2\pi}\nonumber=\frac{1}{\pi}\cos^{-1}\left(\frac{m_{\rm th}-m_X^0}{\delta m}\right)\,.  
\end{align}

\bibliographystyle{JHEP}
\bibliography{biblio}

@article{Hui:2016ltb,
    author = "Hui, Lam and Ostriker, Jeremiah P. and Tremaine, Scott and Witten, Edward",
    title = "{Ultralight scalars as cosmological dark matter}",
    eprint = "1610.08297",
    archivePrefix = "arXiv",
    primaryClass = "astro-ph.CO",
    doi = "10.1103/PhysRevD.95.043541",
    journal = "Phys. Rev. D",
    volume = "95",
    number = "4",
    pages = "043541",
    year = "2017"
}

@article{Graham:2021ggy,
    author = "Graham, Matt and Hearty, Christopher and Williams, Mike",
    title = "{Searches for Dark Photons at Accelerators}",
    eprint = "2104.10280",
    archivePrefix = "arXiv",
    primaryClass = "hep-ph",
    doi = "10.1146/annurev-nucl-110320-051823",
    journal = "Ann. Rev. Nucl. Part. Sci.",
    volume = "71",
    pages = "37--58",
    year = "2021"
}

@article{Martini:2018hhh,
    author = "Martini, Alberto",
    editor = "Bruno, G. E. and Chiodini, G. and Creanza, D. M. and Colangelo, P. and Corian\`o, C. and De Fazio, F. and Nappi, E.",
    title = "{The Belle II experiment: Status and prospects}",
    doi = "10.1051/epjconf/201819200028",
    journal = "EPJ Web Conf.",
    volume = "192",
    pages = "00028",
    year = "2018"
}

@article{Franzoso:2021yjn,
    author = "Franzoso, E.",
    collaboration = "LHCb RICH",
    title = "{The upgrade and performance of the LHCb RICH detector}",
    doi = "10.1393/ncc/i2021-21046-1",
    journal = "Nuovo Cim. C",
    volume = "44",
    number = "2-3",
    pages = "46",
    year = "2021"
}

@article{LHCb:2019vmc,
    author = "Aaij, Roel and others",
    collaboration = "LHCb",
    title = "{Search for $A'\to\mu^+\mu^-$ Decays}",
    eprint = "1910.06926",
    archivePrefix = "arXiv",
    primaryClass = "hep-ex",
    reportNumber = "LHCb-PAPER-2019-031, CERN-EP-2019-212",
    doi = "10.1103/PhysRevLett.124.041801",
    journal = "Phys. Rev. Lett.",
    volume = "124",
    number = "4",
    pages = "041801",
    year = "2020"
}

@article{Belle-II:2024wtd,
    author = "Adachi, I. and others",
    collaboration = "Belle-II",
    title = "{Search for a {\ensuremath{\mu}}+{\ensuremath{\mu}}- resonance in four-muon final states at Belle II}",
    eprint = "2403.02841",
    archivePrefix = "arXiv",
    primaryClass = "hep-ex",
    reportNumber = "2024-007, 2023-55",
    doi = "10.1103/PhysRevD.109.112015",
    journal = "Phys. Rev. D",
    volume = "109",
    number = "11",
    pages = "112015",
    year = "2024"
}

@online{lhcb,
    author = "Jakob Salfeld-Nebgen",
    title="{Latest Results on Dark Photons at the LHC}",
    url = "https://indico.cern.ch/event/900402/contributions/3838434/attachments/2028445/3394353/DarkPhotonLHCDMWG.pdf"
}

@article{Bauer:2018onh,
    author = "Bauer, Martin and Foldenauer, Patrick and Jaeckel, Joerg",
    title = "{Hunting All the Hidden Photons}",
    eprint = "1803.05466",
    archivePrefix = "arXiv",
    primaryClass = "hep-ph",
    doi = "10.1007/JHEP07(2018)094",
    journal = "JHEP",
    volume = "07",
    pages = "094",
    year = "2018"
}

@article{SHiP:2021nfo,
    author = "Ahdida, C. and others",
    collaboration = "SHiP",
    title = "{The SHiP experiment at the proposed CERN SPS Beam Dump Facility}",
    eprint = "2112.01487",
    archivePrefix = "arXiv",
    primaryClass = "physics.ins-det",
    doi = "10.1140/epjc/s10052-022-10346-5",
    journal = "Eur. Phys. J. C",
    volume = "82",
    number = "5",
    pages = "486",
    year = "2022"
}

@article{SHiP:2025ows,
    author = "Albanese, R. and others",
    collaboration = "SHiP, HI-ECN3 Project Team",
    title = "{SHiP experiment at the SPS Beam Dump Facility}",
    eprint = "2504.06692",
    archivePrefix = "arXiv",
    primaryClass = "hep-ex",
    month = "4",
    year = "2025"
}

@article{Antypas:2022asj,
    author = "Antypas, D. and others",
    title = "{New Horizons: Scalar and Vector Ultralight Dark Matter}",
    eprint = "2203.14915",
    archivePrefix = "arXiv",
    primaryClass = "hep-ex",
    reportNumber = "FERMILAB-PUB-22-262-AD-PPD-T",
    month = "3",
    year = "2022"
}

@article{Cogollo:2024fmq,
    author = "Cogollo, D. and Oviedo-Torres, Y. M. and Queiroz, Farinaldo S. and Villamizar, Yoxara and Zamora-Saa, J.",
    title = "{Search for sub-GeV scalars in $e^+e^-$ collisions}",
    eprint = "2412.19893",
    archivePrefix = "arXiv",
    primaryClass = "hep-ph",
    doi = "10.1140/epjc/s10052-025-15094-w",
    journal = "Eur. Phys. J. C",
    volume = "85",
    number = "12",
    pages = "1404",
    year = "2025"
}

@article{VanTilburg:2015oza,
    author = "Van Tilburg, Ken and Leefer, Nathan and Bougas, Lykourgos and Budker, Dmitry",
    title = "{Search for ultralight scalar dark matter with atomic spectroscopy}",
    eprint = "1503.06886",
    archivePrefix = "arXiv",
    primaryClass = "physics.atom-ph",
    doi = "10.1103/PhysRevLett.115.011802",
    journal = "Phys. Rev. Lett.",
    volume = "115",
    number = "1",
    pages = "011802",
    year = "2015"
}

@article{Hees:2016gop,
    author = "Hees, A. and Gu\'ena, J. and Abgrall, M. and Bize, S. and Wolf, P.",
    title = "{Searching for an oscillating massive scalar field as a dark matter candidate using atomic hyperfine frequency comparisons}",
    eprint = "1604.08514",
    archivePrefix = "arXiv",
    primaryClass = "gr-qc",
    doi = "10.1103/PhysRevLett.117.061301",
    journal = "Phys. Rev. Lett.",
    volume = "117",
    number = "6",
    pages = "061301",
    year = "2016"
}

@article{BACON:2020ubh,
    author = "Beloy, Kyle and others",
    collaboration = "BACON",
    title = "{Frequency ratio measurements at 18-digit accuracy using an optical clock network}",
    eprint = "2005.14694",
    archivePrefix = "arXiv",
    primaryClass = "physics.atom-ph",
    doi = "10.1038/s41586-021-03253-4",
    journal = "Nature",
    volume = "591",
    number = "7851",
    pages = "564--569",
    year = "2021"
}

@article{Kennedy:2020bac,
    author = "Kennedy, Colin J. and Oelker, Eric and Robinson, John M. and Bothwell, Tobias and Kedar, Dhruv and Milner, William R. and Marti, G. Edward and Derevianko, Andrei and Ye, Jun",
    title = "{Precision Metrology Meets Cosmology: Improved Constraints on Ultralight Dark Matter from Atom-Cavity Frequency Comparisons}",
    eprint = "2008.08773",
    archivePrefix = "arXiv",
    primaryClass = "physics.atom-ph",
    doi = "10.1103/PhysRevLett.125.201302",
    journal = "Phys. Rev. Lett.",
    volume = "125",
    number = "20",
    pages = "201302",
    year = "2020"
}

@article{Touboul:2022yrw,
    author = "Touboul, Pierre and others",
    title = "{Result of the MICROSCOPE weak equivalence principle test}",
    eprint = "2209.15488",
    archivePrefix = "arXiv",
    primaryClass = "gr-qc",
    doi = "10.1088/1361-6382/ac84be",
    journal = "Class. Quant. Grav.",
    volume = "39",
    number = "20",
    pages = "204009",
    year = "2022"
}

@inproceedings{Craik:2022riw,
    author = "Craik, Daniel and Ilten, Phil and Johnson, Daniel and Williams, Mike",
    title = "{LHCb future dark-sector sensitivity projections for Snowmass 2021}",
    booktitle = "{Snowmass 2021}",
    eprint = "2203.07048",
    archivePrefix = "arXiv",
    primaryClass = "hep-ph",
    month = "3",
    year = "2022"
}

@article{Belle-II:2018jsg,
    author = "Altmannshofer, W. and others",
    editor = "Kou, E. and Urquijo, P.",
    collaboration = "Belle-II",
    title = "{The Belle II Physics Book}",
    eprint = "1808.10567",
    archivePrefix = "arXiv",
    primaryClass = "hep-ex",
    reportNumber = "KEK Preprint 2018-27, BELLE2-PUB-PH-2018-001, FERMILAB-PUB-18-398-T, JLAB-THY-18-2780, INT-PUB-18-047, UWThPh 2018-26",
    doi = "10.1093/ptep/ptz106",
    journal = "PTEP",
    volume = "2019",
    number = "12",
    pages = "123C01",
    year = "2019",
    note = "[Erratum: PTEP 2020, 029201 (2020)]"
}

@article{Chakraborti:2021hfm,
    author = "Chakraborti, Sreemanti and Feng, Jonathan L. and Koga, James K. and Valli, Mauro",
    title = "{Gamma factory searches for extremely weakly interacting particles}",
    eprint = "2105.10289",
    archivePrefix = "arXiv",
    primaryClass = "hep-ph",
    reportNumber = "UCI-TR-2021-12",
    doi = "10.1103/PhysRevD.104.055023",
    journal = "Phys. Rev. D",
    volume = "104",
    number = "5",
    pages = "055023",
    year = "2021"
}

@article{Batell:2014mga,
    author = "Batell, Brian and Essig, Rouven and Surujon, Ze'ev",
    title = "{Strong Constraints on Sub-GeV Dark Sectors from SLAC Beam Dump E137}",
    eprint = "1406.2698",
    archivePrefix = "arXiv",
    primaryClass = "hep-ph",
    doi = "10.1103/PhysRevLett.113.171802",
    journal = "Phys. Rev. Lett.",
    volume = "113",
    number = "17",
    pages = "171802",
    year = "2014"
}

@article{Hu:2000ke,
    author = "Hu, Wayne and Barkana, Rennan and Gruzinov, Andrei",
    title = "{Cold and fuzzy dark matter}",
    eprint = "astro-ph/0003365",
    archivePrefix = "arXiv",
    doi = "10.1103/PhysRevLett.85.1158",
    journal = "Phys. Rev. Lett.",
    volume = "85",
    pages = "1158--1161",
    year = "2000"
}

@article{Schelfhout_2021,
   title={Fourier transform detection of weak optical transitions in atoms undergoing cyclic routines},
   volume={118},
   ISSN={1077-3118},
   url={http://dx.doi.org/10.1063/5.0034217},
   DOI={10.1063/5.0034217},
   number={1},
   journal={Applied Physics Letters},
   publisher={AIP Publishing},
   author={Schelfhout, Jesse S. and Toms-Hardman, Lilani D. and McFerran, John J.},
   year={2021},
   month=jan }

@article{Panelli_2020,
   title={Applying the matched-filter technique to the search for dark matter transients with networks of quantum sensors},
   volume={7},
   ISSN={2196-0763},
   url={http://dx.doi.org/10.1140/epjqt/s40507-020-00081-9},
   DOI={10.1140/epjqt/s40507-020-00081-9},
   number={1},
   journal={EPJ Quantum Technology},
   publisher={Springer Science and Business Media LLC},
   author={Panelli, Guglielmo and Roberts, Benjamin M. and Derevianko, Andrei},
   year={2020},
   month=feb }

@article{Stadnik:2014tta,
    author = "Stadnik, Y. V. and Flambaum, V. V.",
    title = "{Searching for dark matter and variation of fundamental constants with laser and maser interferometry}",
    eprint = "1412.7801",
    archivePrefix = "arXiv",
    primaryClass = "hep-ph",
    doi = "10.1103/PhysRevLett.114.161301",
    journal = "Phys. Rev. Lett.",
    volume = "114",
    pages = "161301",
    year = "2015"
}

@article{Safronova:2017xyt,
    author = "Safronova, M. S. and Budker, D. and DeMille, D. and Kimball, Derek F. Jackson and Derevianko, A. and Clark, C. W.",
    title = "{Search for New Physics with Atoms and Molecules}",
    eprint = "1710.01833",
    archivePrefix = "arXiv",
    primaryClass = "physics.atom-ph",
    doi = "10.1103/RevModPhys.90.025008",
    journal = "Rev. Mod. Phys.",
    volume = "90",
    number = "2",
    pages = "025008",
    year = "2018"
}

@article{Stadnik:2016zkf,
    author = "Stadnik, Y. V. and Flambaum, V. V.",
    title = "{Improved limits on interactions of low-mass spin-0 dark matter from atomic clock spectroscopy}",
    eprint = "1605.04028",
    archivePrefix = "arXiv",
    primaryClass = "physics.atom-ph",
    doi = "10.1103/PhysRevA.94.022111",
    journal = "Phys. Rev. A",
    volume = "94",
    number = "2",
    pages = "022111",
    year = "2016"
}

@article{Arvanitaki:2014faa,
    author = "Arvanitaki, Asimina and Huang, Junwu and Van Tilburg, Ken",
    title = "{Searching for dilaton dark matter with atomic clocks}",
    eprint = "1405.2925",
    archivePrefix = "arXiv",
    primaryClass = "hep-ph",
    doi = "10.1103/PhysRevD.91.015015",
    journal = "Phys. Rev. D",
    volume = "91",
    number = "1",
    pages = "015015",
    year = "2015"
}

@article{Olive:2001vz,
    author = "Olive, Keith A. and Pospelov, Maxim",
    title = "{Evolution of the fine structure constant driven by dark matter and the cosmological constant}",
    eprint = "hep-ph/0110377",
    archivePrefix = "arXiv",
    reportNumber = "UMN-TH-2028-01, TPI-MINN-01-46, MCGILL-01-21",
    doi = "10.1103/PhysRevD.65.085044",
    journal = "Phys. Rev. D",
    volume = "65",
    pages = "085044",
    year = "2002"
}

@article{Schive:2014dra,
    author = "Schive, Hsi-Yu and Chiueh, Tzihong and Broadhurst, Tom",
    title = "{Cosmic Structure as the Quantum Interference of a Coherent Dark Wave}",
    eprint = "1406.6586",
    archivePrefix = "arXiv",
    primaryClass = "astro-ph.GA",
    doi = "10.1038/nphys2996",
    journal = "Nature Phys.",
    volume = "10",
    pages = "496--499",
    year = "2014"
}

@article{Sibiryakov:2020eir,
    author = "Sibiryakov, Sergey and S{\o}rensen, Philip and Yu, Tien-Tien",
    title = "{BBN constraints on universally-coupled ultralight scalar dark matter}",
    eprint = "2006.04820",
    archivePrefix = "arXiv",
    primaryClass = "hep-ph",
    reportNumber = "DESY-19-234, CERN-TH-2020-091, INR-TH-2020-001",
    doi = "10.1007/JHEP12(2020)075",
    journal = "JHEP",
    volume = "12",
    pages = "075",
    year = "2020"
}

@article{Touboul:2017grn,
    author = "Touboul, Pierre and others",
    title = "{MICROSCOPE Mission: First Results of a Space Test of the Equivalence Principle}",
    eprint = "1712.01176",
    archivePrefix = "arXiv",
    primaryClass = "astro-ph.IM",
    doi = "10.1103/PhysRevLett.119.231101",
    journal = "Phys. Rev. Lett.",
    volume = "119",
    number = "23",
    pages = "231101",
    year = "2017"
}

@article{Blum:2014vsa,
    author = "Blum, Kfir and D'Agnolo, Raffaele Tito and Lisanti, Mariangela and Safdi, Benjamin R.",
    title = "{Constraining Axion Dark Matter with Big Bang Nucleosynthesis}",
    eprint = "1401.6460",
    archivePrefix = "arXiv",
    primaryClass = "hep-ph",
    doi = "10.1016/j.physletb.2014.07.059",
    journal = "Phys. Lett. B",
    volume = "737",
    pages = "30--33",
    year = "2014"
}

@article{Batell:2009yf,
    author = "Batell, Brian and Pospelov, Maxim and Ritz, Adam",
    title = "{Probing a Secluded U(1) at B-factories}",
    eprint = "0903.0363",
    archivePrefix = "arXiv",
    primaryClass = "hep-ph",
    doi = "10.1103/PhysRevD.79.115008",
    journal = "Phys. Rev. D",
    volume = "79",
    pages = "115008",
    year = "2009"
}

@article{Bauer:2024yow,
    author = "Bauer, Martin and Chakraborti, Sreemanti",
    title = "{Validity of bounds on light axions for f{\ensuremath{\lesssim}}1013{\,}{\,}GeV}",
    eprint = "2408.06408",
    archivePrefix = "arXiv",
    primaryClass = "hep-ph",
    reportNumber = "IPPP/24/55",
    doi = "10.1103/l94k-l152",
    journal = "Phys. Rev. D",
    volume = "112",
    number = "10",
    pages = "103019",
    year = "2025"
}

@article{Bauer:2024hfv,
    author = "Bauer, Martin and Chakraborti, Sreemanti and Rostagni, Guillaume",
    title = "{Axion bounds from quantum technology}",
    eprint = "2408.06412",
    archivePrefix = "arXiv",
    primaryClass = "hep-ph",
    reportNumber = "IPPP/24/53, IPPP/24/53",
    doi = "10.1007/JHEP05(2025)023",
    journal = "JHEP",
    volume = "05",
    pages = "023",
    year = "2025"
}

@article{Beadle:2023flm,
    author = "Beadle, Carl and Ellis, Sebastian A. R. and Quevillon, J{\'e}r{\'e}mie and Hoa Vuong, Pham Ngoc",
    title = "{Quadratic coupling of the axion to photons}",
    eprint = "2307.10362",
    archivePrefix = "arXiv",
    primaryClass = "hep-ph",
    reportNumber = "CERN-TH-2023-136, DESY-23-097",
    doi = "10.1103/PhysRevD.110.035019",
    journal = "Phys. Rev. D",
    volume = "110",
    number = "3",
    pages = "035019",
    year = "2024"
}

@article{Banerjee:2022sqg,
    author = "Banerjee, Abhishek and Perez, Gilad and Safronova, Marianna and Savoray, Inbar and Shalit, Aviv",
    title = "{The phenomenology of quadratically coupled ultra light dark matter}",
    eprint = "2211.05174",
    archivePrefix = "arXiv",
    primaryClass = "hep-ph",
    doi = "10.1007/JHEP10(2023)042",
    journal = "JHEP",
    volume = "10",
    pages = "042",
    year = "2023"
}

@article{Kim:2023pvt,
    author = "Kim, Hyungjin and Lenoci, Alessandro and Perez, Gilad and Ratzinger, Wolfram",
    title = "{Probing an ultralight QCD axion with electromagnetic quadratic interaction}",
    eprint = "2307.14962",
    archivePrefix = "arXiv",
    primaryClass = "hep-ph",
    reportNumber = "DESY-23-110",
    doi = "10.1103/PhysRevD.109.015030",
    journal = "Phys. Rev. D",
    volume = "109",
    number = "1",
    pages = "015030",
    year = "2024"
}

@article{Hook:2017psm,
    author = "Hook, Anson and Huang, Junwu",
    title = "{Probing axions with neutron star inspirals and other stellar processes}",
    eprint = "1708.08464",
    archivePrefix = "arXiv",
    primaryClass = "hep-ph",
    doi = "10.1007/JHEP06(2018)036",
    journal = "JHEP",
    volume = "06",
    pages = "036",
    year = "2018"
}

@article{Balkin:2020dsr,
    author = "Balkin, Reuven and Serra, Javi and Springmann, Konstantin and Weiler, Andreas",
    title = "{The QCD axion at finite density}",
    eprint = "2003.04903",
    archivePrefix = "arXiv",
    primaryClass = "hep-ph",
    reportNumber = "TUM-HEP-1255/20, TUM-HEP-1255/20",
    doi = "10.1007/JHEP07(2020)221",
    journal = "JHEP",
    volume = "07",
    pages = "221",
    year = "2020"
}

@article{Ferber:2015jzj,
    author = "Ferber, Torben",
    title = "{Towards First Physics at Belle II}",
    doi = "10.5506/APhysPolB.46.2285",
    journal = "Acta Phys. Polon. B",
    volume = "46",
    number = "11",
    pages = "2285",
    year = "2015"
}

\end{document}